\documentclass[twocolumn]{aastex63}
\usepackage[]{units}
\usepackage[update,prepend]{epstopdf}
\usepackage{graphicx}
\usepackage{amssymb}
\usepackage{amsmath}
\usepackage{threeparttable}
\usepackage{hyperref}
\usepackage{color}

\usepackage{mathtools}

\newcommand{\wise}{{\it WISE}}

\newcommand{\kms} {\,km\,s$^{-1}$}
\newcommand{\kmskpc} {\,km\,s$^{-1}$\,kpc$^{-1}$}

\newcommand{\Msun}{\,M$_\odot$}
\newcommand{\Rsun}{\,R$_\odot$}

\mathchardef\mhyphen="2D
\shorttitle{Lifetime of short-period binaries}
\shortauthors{Hwang et al.}

\graphicspath {{figs/} }

\begin{document}

\title{Lifetime of short-period binaries measured from their Galactic kinematics}

\author[0000-0003-4250-4437]{Hsiang-Chih Hwang}
\affiliation{Department of Physics \& Astronomy, Johns Hopkins University, Baltimore, MD 21218, USA}

\author[0000-0001-6100-6869]{Nadia L. Zakamska}
\affiliation{Department of Physics \& Astronomy, Johns Hopkins University, Baltimore, MD 21218, USA}

\begin{abstract}
As a significant fraction of stars are in multiple systems, binaries play a crucial role in stellar evolution. Among short-period ($<$1 day) binary characteristics, age remains one of the most difficult to measure. In this paper, we constrain the lifetime of short-period binaries through their kinematics. With the kinematic information from Gaia Data Release 2 and light curves from {\it Wide-field Infrared Survey Explorer} (WISE), we investigate the eclipsing binary fraction as a function of kinematics for a volume-limited main-sequence sample. We find that the eclipsing binary fraction peaks at a tangential velocity of $10^{1.3-1.6}$\kms, and decreases towards both low and high velocity end. This implies that thick disk and halo stars have eclipsing binary fraction $\gtrsim 10$ times smaller than the thin-disk stars. This is further supported by the dependence of eclipsing binary fraction on the Galactic latitude. Using Galactic models, we show that our results are inconsistent with any known dependence of binary fraction on metallicity. Instead, our best-fit models suggest that the formation of these short-period binaries is delayed by $0.6$-3\,Gyr, and the disappearing time is less than the age of the thick disk. The delayed formation time of $\gtrsim0.6$\,Gyr implies that these short-period main-sequence binaries cannot be formed by pre-main sequence interaction and the Kozai-Lidov mechanism alone, and suggests that magnetic braking plays a key role in their formation. Because the main-sequence lifetime of our sample is longer than 14\,Gyr, if the disappearance of short-period binaries in the old population is due to their finite lifetime, our results imply that most ($\gtrsim90$\%) short-period binaries in our sample merge during their main-sequence stage.

\end{abstract}
\keywords{binaries: close ---  binaries: eclipsing --- stars: kinematics and dynamics}

\section{Introduction}










Binaries are at the core of many exotic astronomical events in the Universe, including classical novae \citep{Warner1995}, red novae \citep{Tylenda2011}, type Ia supernovae \citep{Whelan1973, Iben1984, Webbink1984}, short gamma-ray bursts \citep{Shibata2006,Fong2013}, binary black hole mergers \citep{Abbott2016}, and kilonovae \citep{Abbott2017, Smartt2017, Cowperthwaite2017}. A significant fraction of all stars are in binary and multiple systems \citep{Duchene2013}. Therefore, binary evolution plays a crucial role in the understanding of the Universe.

All the stellar binaries are once a main-sequence (MS) binary. While thousands of short-period ($<1$\,day) MS binaries have been found, they are not formed with such short separation because the radii of pre-MS stars are larger than the MS stars. In fact, the initial separation of binaries is believed to be $\gtrsim10$\,AU because the radius of an initial hydrostatic stellar core is $\sim5$\,AU \citep{Larson1969} and its fragmentation is unlikely \citep{Bate1998,Bate2011}. Therefore, short-period binaries must have gone through orbital migration to shrink the separation from $>10$\,AU ($>2000$\Rsun) to a few \Rsun. 


Short-period binaries may have experienced several different processes to lose orbital angular momentum. At the pre-MS phase, the energy dissipation due to the interaction with the primordial gas may be able to produce binaries with separations down to $\sim0.1$\,AU \citep{Bate2002, Bate2009, Bate2012}. This process takes place on a free-fall timescale, typically $\sim$Myr, and may be able to explain the formation of pre-MS binary stars with periods $>1$\,day \citep{Mathieu1994, Tohline2002}. 

During the MS phase, if a binary has a distant tertiary companion, the angular momentum of the inner binary can exchange with the outer tertiary companion, the so-called Kozai-Lidov mechanism \citep{Kozai1962, Lidov1962}. The inner binary separation can be reduced to a few stellar radii at the pericenter passages due to the high eccentricity excited by the Kozai-Lidov mechanism, and at that point the tidal friction is able to remove the angular momentum and shrink the orbit \citep{Harrington1968,Kiseleva1998, Eggleton2001,Fabrycky2007}. This process is often referred to as Kozai cycles with tidal friction (KCTF). If higher-order effects are taken into accounts, for example eccentric outer orbit and post-Newtonian effects, three-body interactions are more complicated and even chaotic \citep{Naoz2013, Naoz2013a, Naoz2016}. The Kozai-Lidov mechanism is supported by observations that a large fraction of MS close binaries are in triple systems \citep{Tokovinin1997, Pribulla2006, Tokovinin2006, Rucinski2007a}. Furthermore, \cite{Borkovits2016} find that the distribution of mutual inclination between the inner binaries and the outer tertiaries shows a peak at $\sim40^{\circ}$, consistent with the prediction from KCTF \citep{Fabrycky2007}, although the other observed peak at $\sim0^{\circ}$ is not expected. Depending on the configuration of the triple stars and the initial conditions, KCTF may operate on a wide range of timescales, from $\lesssim0.1$\,Gyr to several Gyr \citep{Fabrycky2007, Perets2009}, until the orbits of the inner binaries are circularized.

If the binary separation is close enough (periods$\lesssim5$\,days), magnetic winds become important in extracting angular momentum of binaries. Specifically, stars with masses $\lesssim 1.3$\Msun\ possess subphotospheric convection zones that generate magnetic winds which take away the (rotational) angular momentum of the star. Due to the synchronization between the rotational and orbital periods in short-period binaries, the loss of angular momentum shrinks the orbit. Over a timescale of a few Gyr, magnetic winds are able to bring binaries to the contact phase \citep{Stepien1995,Yakut2008, Van2019, Van2019a}. 




The relative contribution of each process to the formation of short-period ($<1$\,day) MS binaries is not yet clear. In particular, neither pre-MS disk migration nor KCTF during the MS can produce solar-type binaries within orbital periods $<1$\,day. During the pre-MS phase, solar-mass stars accrete most of their mass while slightly enlarged (a few \Rsun), and so disk migration might be able to produce close solar-type binaries down to periods of $\sim1$\,day, below which they would have merged \citep{Hosokawa2009}. Hydrodynamical simulations show that pre-MS interaction is not able to produce binaries with separations $\lesssim 0.1$\,AU, although it can be due to the simulation resolution limit \citep{Bate2002, Bate2009, Bate2012}. The Kozai-Lidov mechanism encounters the difficulty that the majority of binaries with periods $<10$\,days have circular orbits \citep{Latham2002, El-Badry2018}, in which case the Kozai-Lidov mechanism is not effective. Probably only magnetic braking is able to bring solar-type binaries to orbital orbital periods $<1$\,day, but magnetic braking requires small initial period ($\lesssim 5$\,days) to be efficient.

If the ages of short-period binaries could be measured directly, ages could be used to constrain the mechanisms responsible for the orbital migration. By comparing with the MS lifetime, we can determine whether short-period binaries can survive for the entire MS lifetime. If the short-period binaries destruct or disappear at a particular age or at a particular stage of the stellar evolution, then we can constrain or identify the destruction process. 


Age is notoriously difficult to measure for single stars. Furthermore, such methods, including isochrone fitting, stellar rotation, and chromosphere activity, are not valid anymore for short-period binaries because they may have undergone binary interaction and mass transfer. Kinematics is among the few reliable ways to probe the age of short-period binaries. The age-velocity dispersion relation has been well established for a variety of MS stars \citep{Dehnen1998, Nordstrom2004, Reid2009, Sharma2014a}. For disk stars, this relation may be the consequence of kinematic heating processes from giant molecular clouds, transient spiral arms, bars, and flyby satellite galaxies. Kinematics can also help separate the thin-disk, thick-disk, and halo stars. Because accelerations experienced by test particles are independent of their masses, we are able to directly compare the kinematics between single stars and binaries, and further infer their ages.

In this paper, we use Gaia Data Release 2 and the light curves from WISE to investigate the kinematics of short-period ($<1$\,day) eclipsing binaries. In Sec.~\ref{sec:selection}, we describe the dataset, our sample selection, and our time-series analysis. In Sec.~\ref{sec:result} we present our primary results of the relation between eclipsing binary fraction and kinematics. In Sec.~\ref{sec:systematics}, we investigate possible systematics and different sample selections. In Sec.~\ref{sec:galactic-model}, we use Galactic models and show that our results can be explained by a finite lifetime of eclipsing binaries. In Sec.~\ref{sec:discussion}, we discuss disk/halo difference, metallicity, and the implication from the lifetime of eclipsing binaries. We summarize in Sec.~\ref{sec:conclusion}.

\section{Sample Selection and Measurements}
\label{sec:selection}

\subsection{The parent Gaia sample}

Our sample is selected from Gaia Data Release 2 (DR2; \citealt{Gaia2016,Gaia2018Brown}). Gaia is an optical all-sky survey which is obtaining photometry and astrometry for stars with magnitudes down to $\sim21$\,mag and radial velocities for select bright stars. Gaia DR2 was released on 25 April 2018, based on data collected between 25 July 2014 and 23 May 2016. In Gaia DR2, broad-filter G-band magnitudes, blue-band BP magnitudes, red-band RP magnitudes, positions, parallaxes, and proper motions are available for $\sim1.33$ billion objects and radial velocities for $\sim7$ million stars, providing an unprecedented dataset on the phase-space distribution of stars in the Milky Way. 


Our query for Gaia DR2 follows the one used in \cite{Gaia2018Babusiaux}. Specifically, the mean flux divided by its error is is larger than 50 for G-band and larger than 20 for BP and RP bands. In Gaia DR2, BP and RP fluxes are not treated with deblending, so we apply a cut on {\tt phot\_bp\_rp\_excess\_factor} to reduce the effect of crowded fields which makes the BP and RP bands unreliable \citep{Evans2018, Arenou2018}. {\tt visibility\_periods\_used}$>8$ is used to ensure that there are sufficient observations for deriving the astrometric solutions \citep{Lindegren2018}, and {\tt parallax\_over\_error}$>10$ is adopted to have well-measured parallaxes. We do not apply explicit cuts on relative proper motion uncertainties because that excludes objects having intrinsically low proper motions, which biases the kinematic results. Instead, we follow \cite{Gaia2018Babusiaux} and use the unit error introduced by \cite{Lindegren2018} to avoid spurious astrometric solutions. Because we obtain the light curves from WISE, we cross-match Gaia DR2 and WISE using the Gaia DR2 cross-match catalog \citep{Marrese2019}. Our Gaia DR2 query is included in the Appendix. 


We compute tangential velocities from proper motions and parallaxes provided by Gaia DR2. We do not use the radial velocities in Gaia DR2 because the radial velocity sample is $\sim100$ times smaller. Furthermore, Gaia DR2 does not report the radial velocities of double-line systems and objects having high radial velocity variations \citep{Katz2019}, which strongly biases the binary selection.

We correct the velocities by removing the solar motion and the differential rotation of the Galactic disk. We adopt the solar motion from \cite{Schonrich2010} where $({\rm U}_\odot,{\rm V}_\odot,{\rm W}_\odot) = (11.1, 12.24, 7.25)$\kms, with the convention that ${\rm U}_\odot,{\rm V}_\odot,{\rm W}_\odot$ are oriented towards the Galactic center, the direction of Galactic rotation, and the north Galactic pole. Our sample is within 500\,pc with a median of 380\,pc, the local shear approximation described by Oort's constants is applicable. We remove the contribution from differential rotation of the Galactic disk using the Oort constants reported from \cite{Bovy2017}: $A=15.3$\kmskpc, $B=-11.9$\kmskpc, $C=-3.2$\kmskpc, $K=-3.3$\kmskpc. While this correction is only valid for disk stars and not for halo stars because halo stars are not rotating with the disk, but since the velocity correction of differential rotation is $<10$\kms, this (incorrect) correction is small for halo stars where the typical velocities are $>100$\kms. With the correction of the solar motion and the differential rotation of the Galactic disk, the tangential velocities ($V_t$) presented in this paper are the tangential components relative to the local Galactic rotation at the star's location.

\subsection{Main sequence selection}

Our MS selection is designed to satisfy several purposes: (1) the binary fraction is a strong function of mass and therefore color \citep{Duchene2013}. Using a narrow color range reduces such mass dependence in our results. (2) On the blue end (BP$-$RP$<0.5$\,mag), there is contamination from pulsating stars like $\delta$ Scuti. (3) Stars leave the MS phase because of stellar evolution, so selecting long-lived (i.e. redder) MS stars helps to interpret kinematic results. (4) Because binaries are brighter than single stars, we aim to construct a volume-limited sample instead of a magnitude-limited sample to avoid systematics. For this reason, we cannot use MS stars that are too red because they are faint and the sample size would be small in a volume-limited sample. 

\begin{figure}
	\centering
	\includegraphics[width=0.9\linewidth]{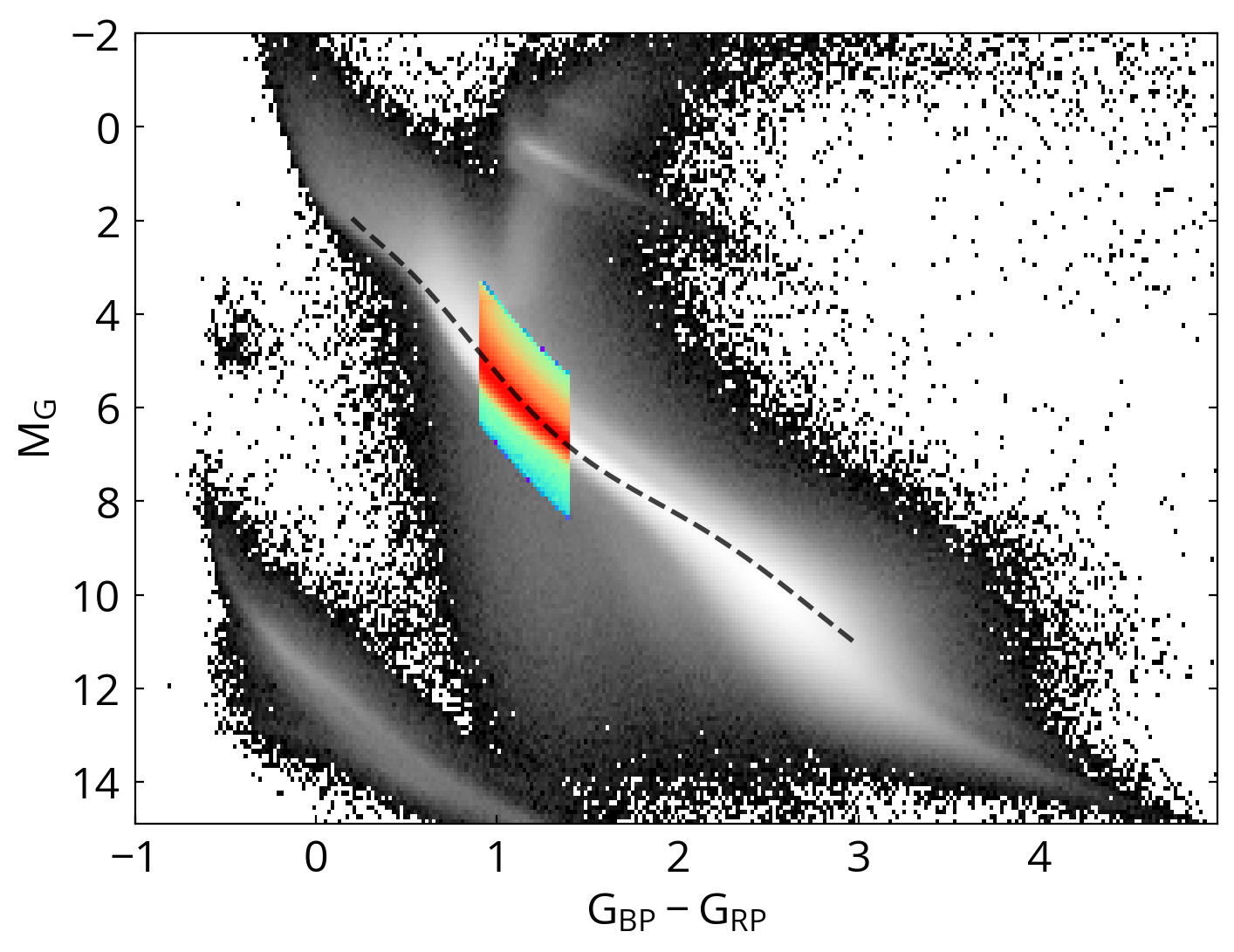}
	\caption{The H-R diagram demonstrating our selection. The x-axis is the Gaia BP$-$RP color, and the y-axis is the Gaia absolute G-band magnitude. The gray scale shows is the stars within 500\,pc, and the color region indicates our main sequence sample. The dashed line is the spline fit of Pleiades.}
	\label{fig:HR}
\end{figure}

To address all these points, we select an MS sample with absolute magnitude offsets $|\Delta \rm G| < 1.5$\,mag and with $0.9<$BP$-$RP$<1.4$, shown as the colored region in Fig.~\ref{fig:HR}. The black dashed line in Fig.~\ref{fig:HR} is the spline fit to Pleiades, following \cite{Hamer2019}. Pleiades is a young, solar-metallicity open cluster with age $10^{8.04}$\,years and [Fe/H]$=-0.01$ \citep{Netopil2016}, and $\Delta \rm G$ is defined as the offset of absolute G magnitudes between the stars and Pleiades at the same BP$-$RP colors, where $\Delta \rm G<0$ means that the star is brighter than Pleiades at the same color. Objects are defined as MS if they have $|\Delta \rm G|<1.5$\,mag. The 1.5 magnitude range is motivated by the fact that we want to include binaries, which are 0.75 mag brighter than single stars assuming equal luminosities and no occultations, and we also want to keep thick-disk and halo stars which are $\lesssim1$\,mag fainter than Pleiades in the color range considered due to their lower metallicities.

The color selection of $0.9<$BP$-$RP$<1.4$ is chosen to avoid the pulsating stars at BP$-$RP$<0.5$, and to include most of the eclipsing binaries concentrated around BP$-$RP$\sim1$ in Fig.~\ref{fig:P-BPRP}. From the PARSEC isochrone \citep{Bressan2012}, the selection of BP$-$RP$=0.9$-$1.4$ has masses ranging from 0.7 to 0.9\Msun, with temperatures 4500-5500\,K, corresponding to late-G and K dwarfs. This selection ensures that their MS lifetime is longer than 14\,Gyr.

\subsection{Eclipsing binary sample from WISE}

\label{sec:time-series}

Because Gaia DR2 has not released the time series and the catalog of eclipsing binaries, we construct the eclipsing binary sample using two other ways. One is from the light curves of WISE, and the second is from the variability information in Gaia DR2. These two samples are complementary: the WISE sample has less contamination, while the Gaia sample is more complete in terms of the sky distribution and is not affected by the limits of period-finding algorithms. 

The WISE eclipsing binary sample is constructed using {\it Wide-field Infrared Survey Explorer} (\wise; \citealt{Wright2010}). Our work requires a large sample and good cadences to recover short-period ($<1$\,day) eclipsing binaries, and WISE serves as an excellent dataset for this purpose. Since WISE is an all-sky survey, it provides a large cross match sample with Gaia. Furthermore, the orbital period of WISE satellite is $\sim1.6$ hours, which is able to recover the MS eclipsing binaries, and this cadence is much better than most of the ground-based surveys. Its W1 (3.4\,\micron) and W2 (4.6\,\micron) bands have been collecting data since AllWISE in 2010 to NeoWISE in 2019, providing a long baseline to study the time series. In main-sequence regions and the color and parallax range of interest, we end up with $\sim1000$ short-period eclipsing binaries in WISE, compared to only a few hundred targets in the Kepler eclipsing binary catalog \citep{Kirk2016} and the Catalina Sky Survey \citep{Drake2014} under the same criteria.

The AllWISE source catalog provides {\tt var\_flg} which is a measure of the probability that an AllWISE source is variable in each WISE filter. Specifically, {\tt var\_flg} is an integer ranging from 0 to 9 such that $\sim10^{\tt-var\_flg}$ is the probability that the observed WISE light curve is drawn from a non-variable population \citep{Hoffman2012}. Therefore, {\tt var\_flg}$=0$ means non-variable and $9$ indicates the highest probability of being variable. Out of the parent Gaia sample, we select targets where {\tt var\_flg}$\ge5$ in W1 band to further analyze their light curves, resulting $\sim20,000$ variable candidates. We do not consider other WISE bands because W2 has worse sensitivity, and W3 and W4 do not have single-epoch exposures in NeoWISE.

We download complete W1 light curves from AllWISE Multiepoch Photometry Table and NEOWISE-R Single Exposure (L1b) Source Table through NASA/IPAC Infrared Science Archive and perform time series analysis for variable candidates where {\tt var\_flg}$\ge5$. The W1 light curves from AllWISE and NeoWISE provide a total baseline of $\sim9$\,years from 2010 to 2019. To ensure the photometric quality of single-epoch exposures, we follow the instruction of \cite{Cutri2011} to adopt the criteria of {\tt saa\_sep}$>0$, {\tt moon\_masked}$=0$, {\tt qi\_fact}$>0.9$ for AllWISE\footnote{http://wise2.ipac.caltech.edu/docs/release/allwise/expsup/sec3\_1.html}, and {\tt saa\_sep}$>0$, {\tt moon\_masked}$=0$, {\tt qi\_fact}$>0.9$, and {\tt qual\_frame}$>0.9$ for NeoWISE\footnote{http://wise2.ipac.caltech.edu/docs/release/neowise/expsup/sec2\_3.html}. AllWISE Multiepoch Photometry Database also contains some redundant photometric measurements, and we further exclude them by matching the source ID (source\_id\_mf) in AllWISE and NeoWISE \footnote{http://wise2.ipac.caltech.edu/docs/release/allwise/expsup/sec3\_2.html}.

We use the periodogram of Multi-Harmonic Analysis of Variance (MHAOV; \citealt{Schwarzenberg-Czerny1996}) to determine the periodicity in the light curves. MHAOV has good performance compared to other period-finding algorithms in terms of the accuracy against magnitude, sampling rates, quoted period, S/N ratios, number of observations, and different variability classes \citep{Graham2013}. We run MHAOV with three harmonics on the WISE variable candidates from $f_{\rm min}=0.1$\,day$^{-1}$ to $f_{\rm max}=20$\,day$^{-1}$ with $\Delta f=1\times10^{-4}$\,day$^{-1}$, or equivalently periods ranging from 0.05\,day (1.2 hour) to 10 days. \cite{Chen2018} also measure the periods for WISE variables, but their minimum period in the periodogram is set to 0.143\,day (3.4 hours). While this minimum period is safer because it is above the classical Nyquist period of 3.2 hours (i.e. two times of the WISE satellite's orbital period), it misses most of short-period MS eclipsing binaries. The Nyquist sampling theorem applies when the sampling is uniform, whereas WISE satellite does not observe targets uniformly due to the size of the field of view and the drift of the satellite's orbital plane \citep{Mainzer2014}. Thus WISE's slightly irregular sampling may help to recover periods below the classical Nyquist limit \citep{VanderPlas2018}. If the aliasing does happen, it results in an aliased peak in the periodogram. Therefore, aliasing only makes the measured period inaccurate but does not affect the fact that such source is a periodic variable. Since our main interest is in selecting short-period ($<1$\,day) eclipsing binaries but not their exact periods, aliasing does not affect our sample selection. Therefore, we adopt a minimum period in the periodogram smaller than the classic Nyquist limit to recover short-period MS eclipsing binaries. A more detailed investigation of the short-period WISE periodic variables will be presented in Petrosky et al. (in prep.). 




For some eclipsing binaries, particularly contact binaries (W UMa binaries), their primary and secondary eclipses have similar depths so period-searching algorithms may not be able to distinguish the primary from secondary eclipses. Therefore, the period-searching algorithm may report a period that is two times smaller than the orbital period. We do not attempt to apply this factor of 2 correction, and we refer to the measured periods from periodograms as `apparent periods', and keep in mind that the apparent periods may be two times smaller than the orbital periods of the binaries.



After time series analysis of the WISE variables, we use the following criteria to select the WISE eclipsing binary sample: (1) the peak in the MHAOV periodogram ($\Theta$ statistics) is larger than 200, meaning that a strong periodic signal is detected in the light curves; (2) there is at least one observation in every 0.05 phase in the phase-folded light curves, ensuring that the light curve is well-sampled; (3) even with the previous two criteria, there is an overdensity in the apparent periods at $\sim0.067$\,day, the orbital period of the WISE satellite. Therefore, we limit our sample to apparent periods $>0.07$\,day to avoid these spurious period measurements. These three criteria result in 2994 periodic variables from the parent Gaia sample (without the MS selection). We inspect their phase-folded light curves and confirm that these criteria provide a robust eclipsing binary sample.  

\begin{figure}
	\centering
	\includegraphics[width=0.9\linewidth]{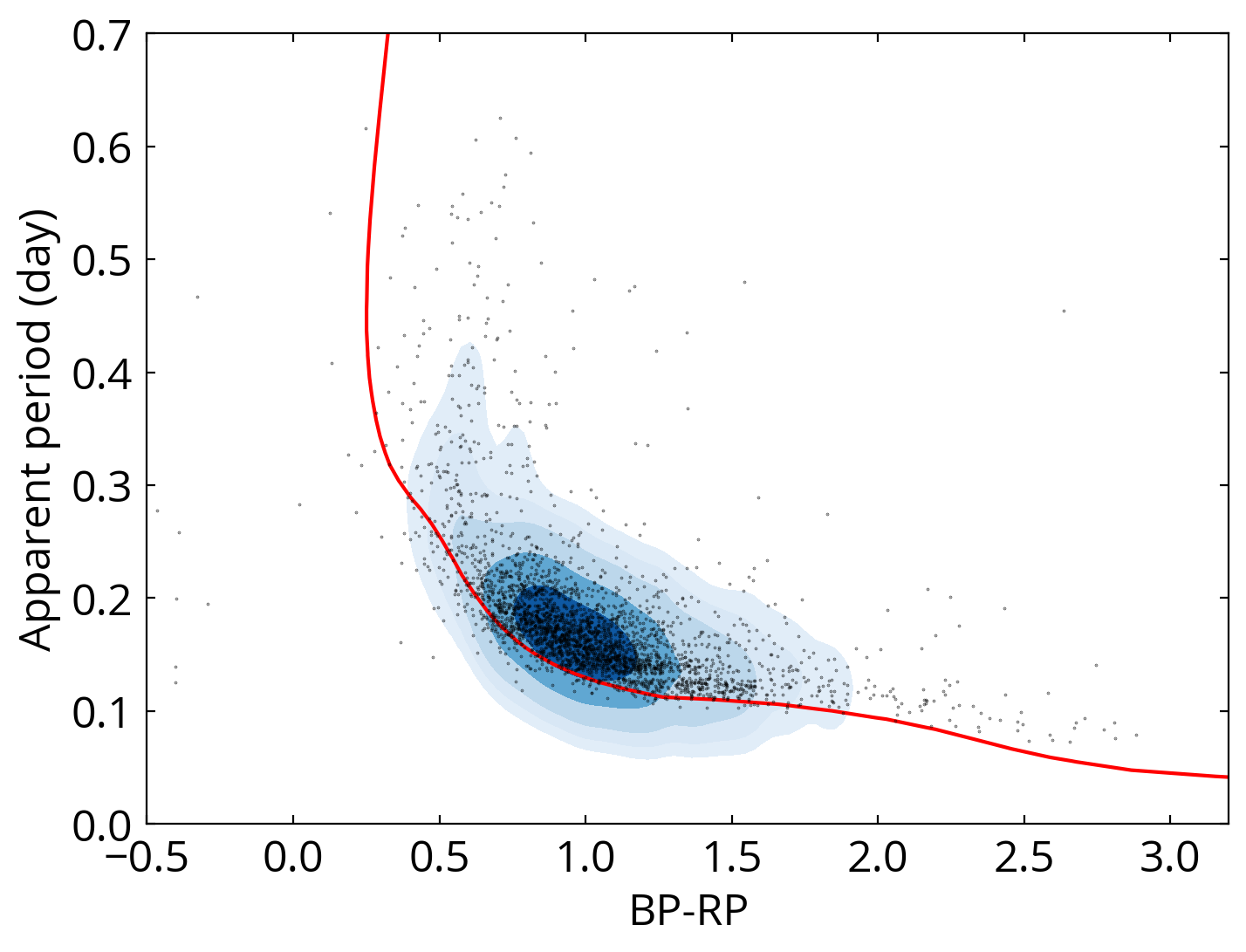}
	\caption{The distribution of apparent periods with respect to Gaia BP$-$RP color for the WISE periodic variables. The y-axis shows the apparent periods reported from periodograms, which typically are half of the orbital periods for short-period eclipsing binaries. The dots are the individual measurements, and the blue background is the Gaussian kernel density estimation where the bandwidths are chosen to present a smooth distribution. The red solid line shows the theoretical apparent periods ($0.5\times$orbital periods) for contact, equal-mass binaries. The WISE variables agree very well with the red solid line, meaning that they are MS eclipsing binaries. }
	\label{fig:P-BPRP}
\end{figure}

Fig.~\ref{fig:P-BPRP} shows the apparent periods of the WISE periodic variables with respect to the Gaia BP$-$RP colors. The red line is the theoretical minimum possible period for contact, equal-mass MS binaries. The red line is derived using PARSEC isochrone \citep{Bressan2012} with an age of 9\,Gyr and solar metallicity. The simple theoretical minimum possible apparent period is computed by $P_{apparent} = 0.5 P_{orbital} = \pi \sqrt{a^3/G(M_1+M_2)}$, where $a$ is the semi-major axis of the binary, $G$ is the gravitational constant, and $M_1$ and $M_2$ are the masses of the stars. We consider equal-mass binaries ($M_1=M_2$) and use the Roche-lobe volume radius $R_L=0.38a$ \citep{Eggleton1983}, where $R_L$ is the volume radius of a star. By definition, the volume radius $R_L$ equals to the radius of an undistorted star which is provided in the PARSEC isochrone. The overall trend of the solid red line in Fig.~\ref{fig:P-BPRP} represents the periods limited by the sizes of stars: bluer, larger stars have larger minimum periods while redder, smaller stars can have smaller periods.

Fig.~\ref{fig:P-BPRP} shows that our period measurements are in excellent agreement with the period limit of contact binaries, meaning that our WISE variable sample, if not all, is dominated by eclipsing binaries.  Fig.~\ref{fig:P-BPRP} also emphasizes the need to search apparent periods below the classical Nyquist limit of 0.13 day, otherwise most of the eclipsing binaries having BP$-$RP$>1$ would be missed. Some narrow gaps in apparent periods at multiples of the WISE's orbital period (0.13\,day and 0.2\,day) can be seen in the black points because the sampling is not sensitive to their periods. While aliasing can potentially downgrade our period accuracy, Fig.~\ref{fig:P-BPRP} shows that our results pass through the classic Nyquist limit at 0.13\,day quite smoothly and recover a large number of low-mass eclipsing binaries below this limit. The blue end (BP$-$RP$\lesssim 0.5$) overlaps the instability strip, so some of them may be $\delta$ Scuti variables \citep{Gaia2018Eyer}. Type-II Cepheids are also located at the blue end (BP$-$RP$\lesssim 0.5$), but their periods typically are longer than 1 day and are not seen in this plot.

\subsection{Eclipsing binary sample from Gaia DR2}
\label{sec:gaia-time-series}


Here we construct another eclipsing binary sample using Gaia DR2 alone. While Gaia DR2 does not release the catalog and the light curves of eclipsing binaries, we can construct an indirect eclipsing binary sample from Gaia DR2. The variability information can be obtained from the photometric errors of Gaia DR2 \citep{Gaia2018Eyer}. The photometric errors are calculated by
\[
{\tt phot\_g\_mean\_flux\_error} = \sigma_G / \sqrt{{\tt phot\_g\_n\_obs}},
\]
where $\sigma_G$ is the standard deviation of the G-band fluxes. When a star passes through the field of view of the Gaia satellite, it goes through 9 astrometric field CCDs where each CCD has one G-band photometric measurement. This $\sigma_G$ is the standard deviation of each CCD photometric measurement, which are obtained within the crossing time of a source over one CCD is $\sim4.4$ seconds. Furthermore, as the Gaia satellite spins with a period of 6 hours, a source passes its two field of views separated by $\sim1.8$ (or 4.2) hours. Therefore, $\sigma_G$ also contains the information on variability on timescales of hours. Depending on the location of the sky, Gaia scans through the same target after several weeks \citep{Evans2018,Riello2018}. In our selection, we require that {\tt visibility\_periods\_used > 8}, ensuring that there are enough visits to derive reliable astrometric solutions but also enough observations to measure photometric variability. In our Gaia sample with the MS cut, the median {\tt visibility\_periods\_used} is 13 and the median {\tt phot\_g\_n\_obs} (number of CCD photometric measurements contributing to G photometry) is 254. 

Based on the photometric errors in the Gaia DR2, we compute $\sigma_G$ and further $f_{G,raw} = \sigma_G/F_G$ for all the sources, where $F_G$ is the mean flux in the G band. We refer to the dimensionless $f_{G,raw}$ as `raw fractional variability' in the G band. While $f_{G,raw}$ contains the information about the variability of stars, it has to be corrected for the magnitude-dependent instrumental errors \citep{Evans2018}. The instrumental fractional variability, $f_{G,inst}$, is computed from the running modes of $f_G$ for our entire sample across the observed G-band magnitudes. Then the instrumentally corrected fractional variability is $f_{G}^2 = f_{G,raw}^2 - f_{G,inst}^2$. In this definition, $f_{G}^2$ may be negative, which means that such star does not have significant variability compared to the instrumental level. 95\% of our MS sample is brighter than 14.8\,mag in G-band, where the instrumental correction is $f_{G,inst}\sim0.8$\%.

\begin{figure}
	\centering
	\includegraphics[width=1.\linewidth]{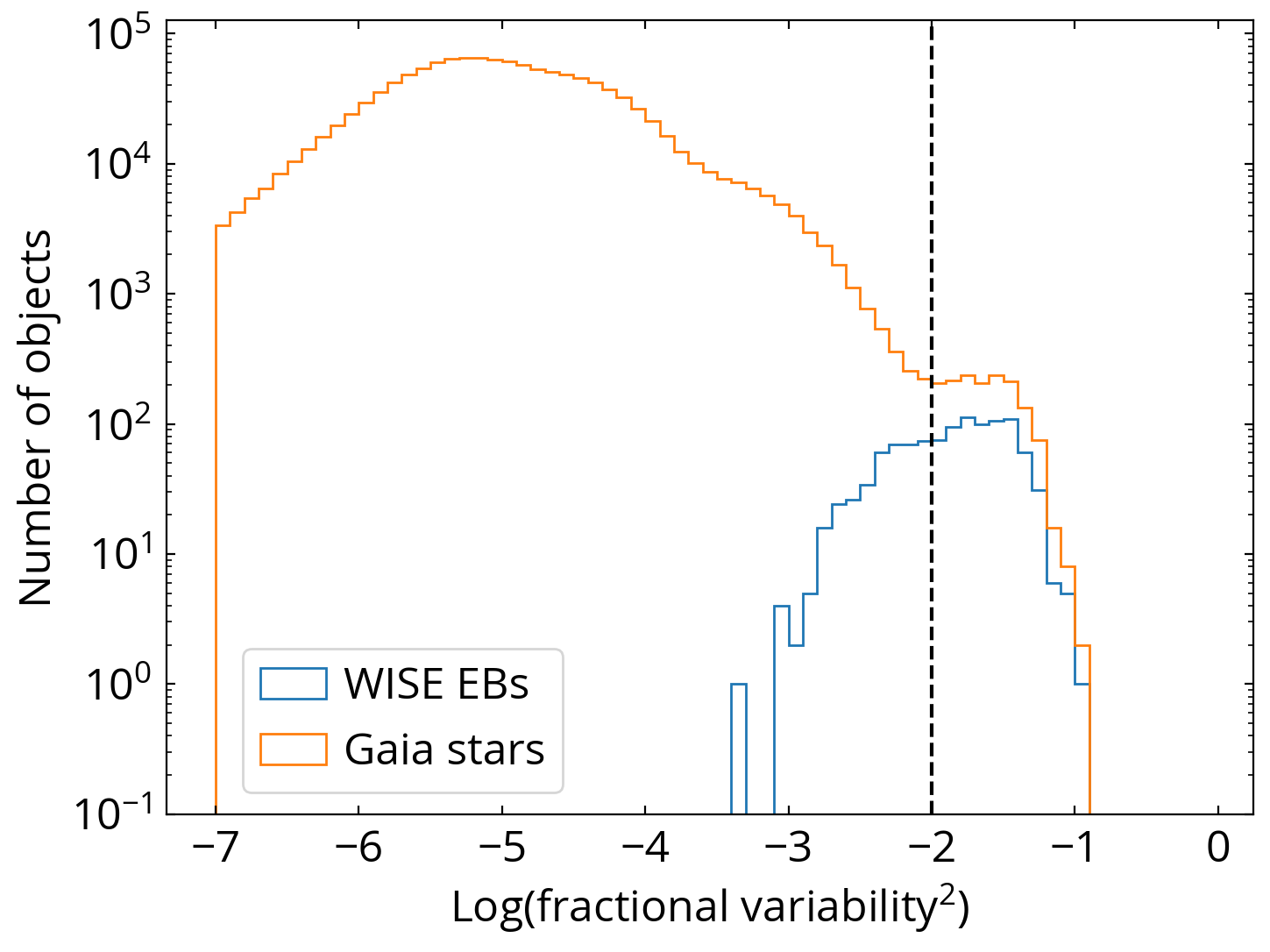}
	\caption{The distribution of fractional variability ($f_G$) for the Gaia main sequence sample and WISE eclipsing binaries. The excess of objects at $\log (f_G^2)\sim -3$ and $>-2$ two physical mechanisms for variability. By comparing with the WISE eclipsing binaries, we select Gaia stars having $\log (f_G^2)>-2$ (dashed line), i.e. variability $>10\%$, as the Gaia eclipsing binary sample. }
	\label{fig:var-metric}
\end{figure}

We use $f_{G}^2$ to identify eclipsing binaries in Gaia DR2. Fig.~\ref{fig:var-metric} shows the distribution of $\log(f_{G}^2)$ for the eclipsing binaries identified from WISE and for all MS stars located in the same region in the H-R diagram. The distribution of MS stars has a small excess at $\log (f_G^2) \sim -3$ and an enhanced tail at $\log (f_G^2)> -2$, suggesting two different origins for variability. By comparing with the WISE eclipsing binaries, we select stars having $\log (f_{G}^2)>-2$ (dashed line) as the eclipsing binary candidates. The excess at $\log (f_{G}^2)\sim -3$ is likely due to stellar rotation and the spots, ellipsoidal variations, and/or (semi-)detached binaries with longer orbital periods (Appendix~\ref{sec:kepler}). Particularly, we find that stars having $-2.5<\log (f_{G}^2)<-2$ are significantly kinematically cooler than other stars, suggesting that they may be young stars where the spots are more active, or young binaries where the orbital periods are larger. A similar method has been used to obtain Gaia variability information to identify RR Lyrae stars \citep{Belokurov2017} and sub-kpc dual quasar candidates \citep{Hwang2019a}.

This selection of eclipsing binaries is based on the assumption that eclipsing binaries are the dominant sources of variability on the MS in the color range considered. Although we do not have the information of periods for eclipsing binaries selected from Gaia DR2, we expect that short-period binaries dominate the sample because systems having shorter orbital periods have a higher probability of being eclipsing systems. Furthermore, eclipsing binaries with orbital periods $>1$\,day tend to be more detached and so vary in brightness only during eclipses, i.e., a smaller duty cycle of variability, which reduces their overall fractional variability. Indeed, in Appendix~\ref{sec:kepler} we show that the majority of eclipsing binaries selected by Gaia fractional variability have orbital periods $<0.5$\,day. While it is still possible that some variability can be due to stellar rotations and flares, we argue that eclipsing binaries still dominate the number. One reason is that our criteria select objects with large variation amplitudes of $f_G^2>0.01$ (i.e. $>10$\%), which is unlikely to be due to spots.



\subsection{Detectability as a function of parallax}
\begin{figure}
	\centering
	\includegraphics[width=1.\linewidth]{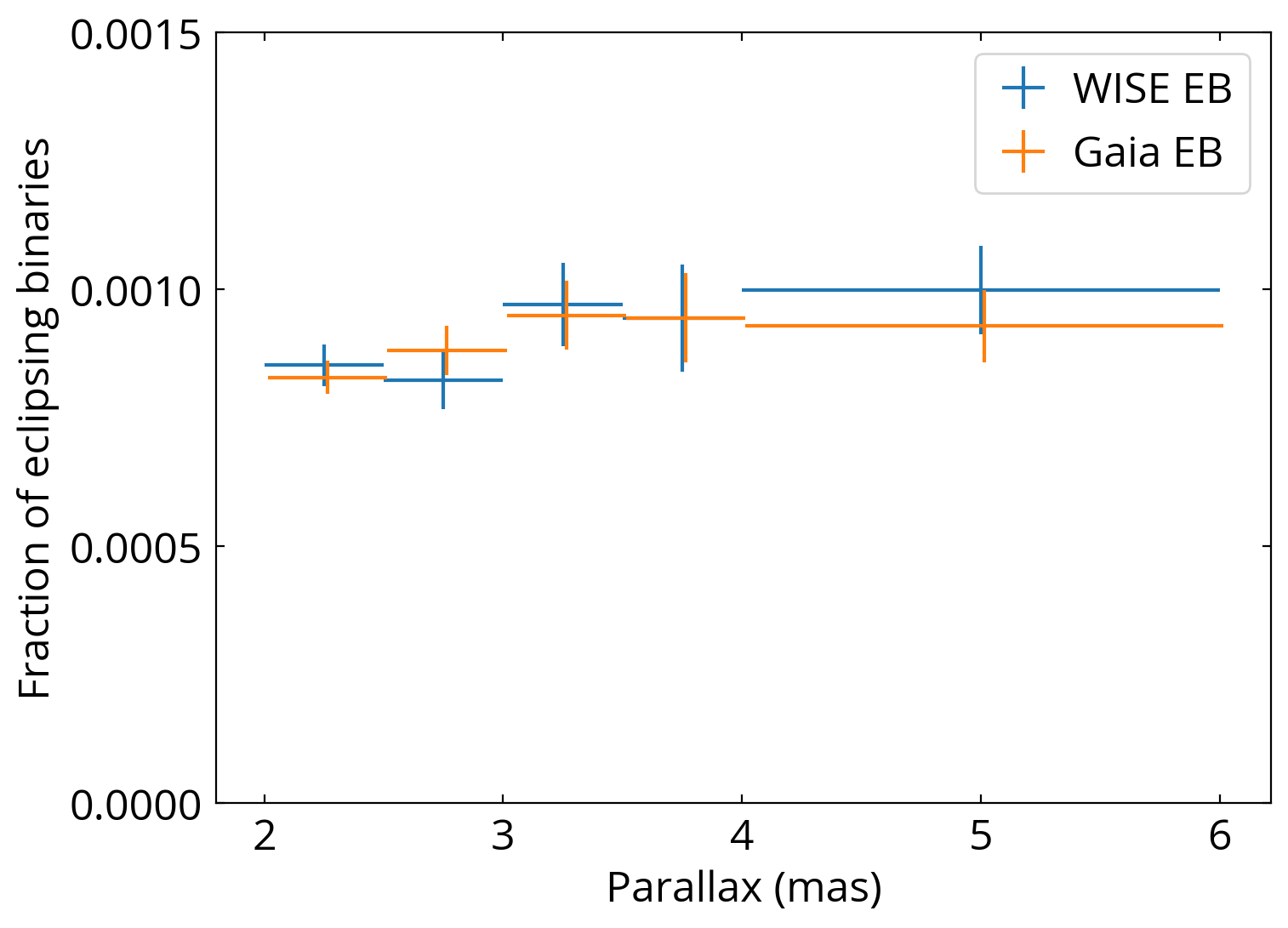}
	\caption{The fraction of eclipsing binaries as a function of parallax for the WISE and Gaia sample. The eclipsing binary fractions remain fairly flat over the entire range of parallax, meaning that we have not reached the limit of detectability of eclipsing binaries. We use parallax $>2$\,mas (i.e. within 500\,pc) as our volume-limited sample.}
	\label{fig:magpar-cut}
\end{figure}

After introducing the main-sequence selection and two eclipsing binary selections, we now determine the parallax (distance) cut to construct a volume-limited eclipsing binaries sample with the MS selection of $|\Delta \rm G| < 1.5$\,mag and $0.9<$BP$-$RP$<1.4$. Fig.~\ref{fig:magpar-cut} shows the fraction of stars that are short-period WISE eclipsing binaries as a function of parallax. If a star is too far so that it is not well detected in a single-exposure in WISE and therefore the periods cannot be well determined, we expect a steep decline in eclipsing binary fraction at a certain parallax. Instead, we see a flat dependence in Fig.~\ref{fig:magpar-cut}, and the slight decrease at 2.5-3 mas parallaxes is likely consistent with the difference in eclipsing binary fraction between the thin disk and the older stars as discussed below.


With the criteria of $|\Delta \rm G| < 1.5$\,mag, $0.9<$BP$-$RP$<1.4$, and parallax $>2$\,mas, we end up with 1081 WISE eclipsing binaries. All of them have apparent periods $<0.5$\,day. We do not correct for dust extinction because the Galactic models we use for comparisons with data includes the effects of extinction. At the limiting distance of 500 pc of our sample, the level of reddening (E(B$-$V)$<0.2$\,mag) is small compared to the color range of our selection, and the level of extinction (A$_{\rm V}<0.8$\,mag) does not affect the completeness of the volume-limited sample in our chosen magnitude range.

Fig.~\ref{fig:magpar-cut} also shows that the Gaia eclipsing binary fraction is consistent with WISE eclipsing binary sample and the detectability stays constant. With the criteria of $|\Delta \rm G| < 1.5$\,mag, $0.9<$BP$-$RP$<1.4$, and parallax $>2$\,mas, we end up with 1545 eclipsing binaries from Gaia DR2.

\subsection{General properties of our eclipsing binary selections}

While we refer to our sample as eclipsing binaries, the variability may not only come from eclipses. The variability can also be the ellipsoidal modulation due to the strongly distorted stars. For WISE eclipsing binaries, we do not attempt to classify  eclipsing binaries into subclasses based on their light curves, but \cite{Paczynski2006} show that eclipsing binaries with periods $<1$\,day consist of mostly contact binaries and some semi-detached binaries, and very few detached binaries. 

Both our WISE and Gaia eclipsing binary selections tend to select binaries with shorter periods because the probability of being eclipsed is higher, and also shorter-period systems are more likely to be contact binaries where photometric variability is stronger. By comparing with the Kepler eclipsing binaries \citep{Kirk2016} in Appendix~\ref{sec:kepler}, we find that the majority of the Gaia eclipsing binaries have orbital periods $<0.5$\,day (or apparent periods $<0.25$\,day). Similarly, Fig.~\ref{fig:P-BPRP} shows that most of our WISE eclipsing binaries have apparent periods $<0.25$\,day, corresponding to orbital periods $<0.5$\,day. Since we are interested in eclipsing binary fraction as a function of kinematics, missing non-eclipsed short-period binaries only affects our sample completeness but does not bias the kinematic result.



\subsection{Summary of the sample selection}


Here we summarize our sample selection. Each of the WISE and Gaia samples has a parent MS sample and an eclipsing binary sample. The parent MS samples have the same selection as their corresponding eclipsing binary samples except without requiring variability or eclipses. For the WISE sample, the selection criteria are:

\begin{enumerate}
	\item {\tt parallax\_over\_error} $>10$.
	\item {\tt phot\_g\_mean\_flux\_over\_error} $> 50$.
	\item {\tt phot\_rp\_mean\_flux\_over\_error} $> 20$.
	\item {\tt phot\_bp\_mean\_flux\_over\_error} $> 20$.
	\item {\tt visibility\_periods\_used} $>8$.
	\item Cuts on {\tt phot\_bp\_rp\_excess\_factor} following \cite{Gaia2018Babusiaux}.
	\item Cuts on unit errors following \cite{Gaia2018Babusiaux}.
	\item {\tt parallax}$>2$\,mas (i.e. within 500\,pc).
	\item MS selection so that the absolute G-band magnitude relative to Pleiades is smaller than 1.5\,mag ($|\Delta G| < 1.5$).
	\item A color selection of $0.9<$BP$-$RP$<1.4$.
	\item {Every object has an AllWISE cross match.}
	\item AllWISE {\tt cc\_flags = 0000}, indicating no spurious signals in WISE images.
	\item For the WISE eclipsing binary selection, we require that the peak in the MHAOV periodograms is larger than 200, at least one observation in every 0.05 phase in the phase-folded light curves, and apparent periods between 0.07 and 0.5\,day.
\end{enumerate}

The Gaia parent MS sample is selected using criteria (1)-(10), and the Gaia eclipsing binaries are selected further using $\log(f_G^2) > -2$. 


\section{Eclipsing binary fractions as a function of kinematics}
\label{sec:result}

\begin{figure*}
	\centering
	\includegraphics[width=.7\linewidth]{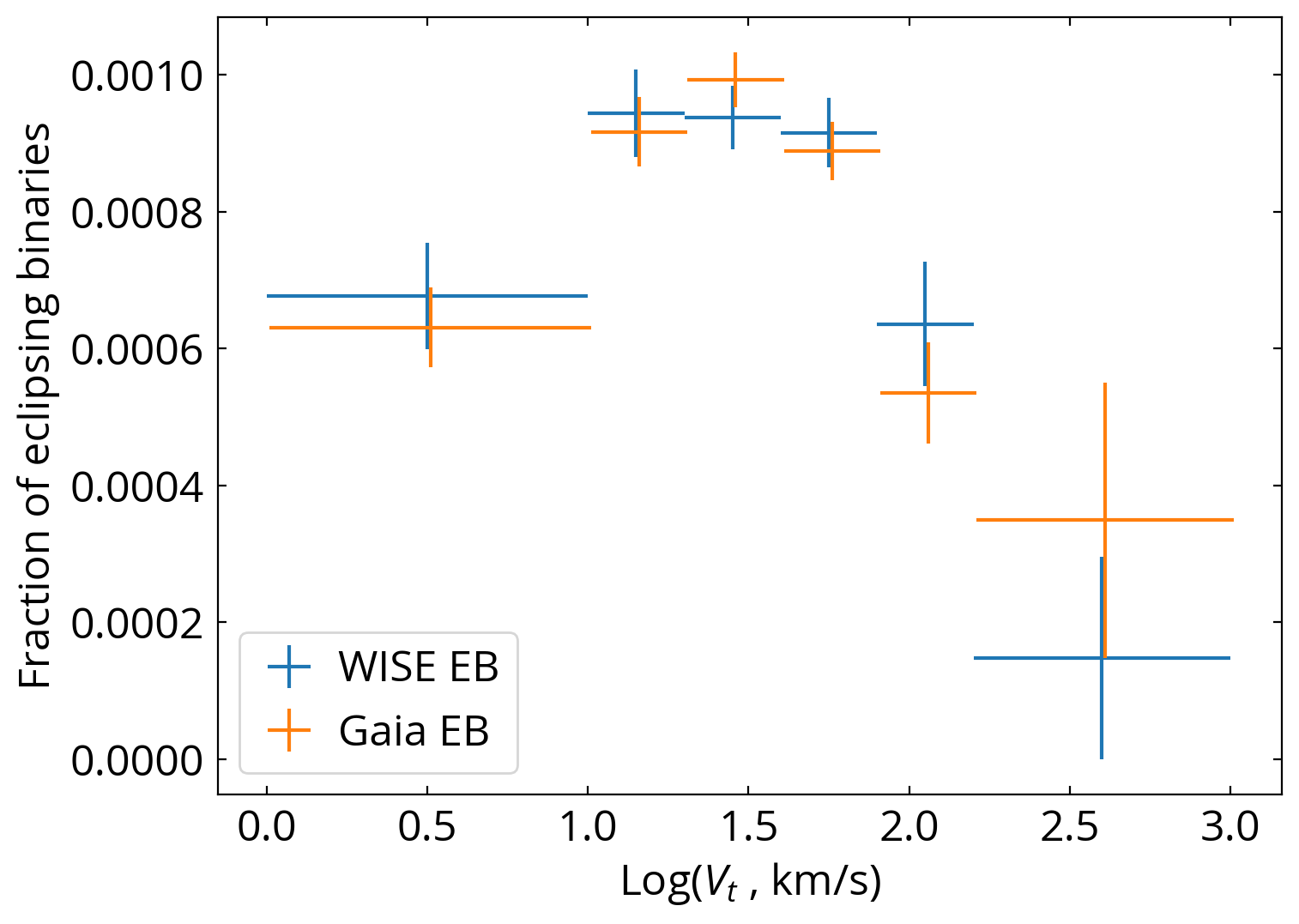}
	\caption{The fraction of short-period eclipsing binaries as a function of tangential velocity. The blue crosses use the eclipsing binaries selected from WISE, and the orange crosses are from Gaia DR2. The horizontal bars indicate the size of the bins, and the vertical bars are errors estimated using Poisson statistics. Both eclipsing binary samples show that the eclipsing binary fraction peaks at tangential velocity of $\sim 10^{1.3-1.6}$\kms, and decreases toward both lower and higher velocities.}
	\label{fig:frac-Vt}
\end{figure*}

Fig.~\ref{fig:frac-Vt} presents the eclipsing binary fraction as a function of tangential velocity in the MS sample. Because WISE eclipsing binaries are easier to identify with more WISE scans and therefore may have different sky distribution as the WISE parent sample, we weigh the WISE result based on the sky distribution of WISE eclipsing binaries. Specifically, we bin the WISE eclipsing binaries by the galactic coordinates with steps of $\Delta l=15$\,deg and $\Delta b=10$\,deg, and assign weights to each bin such that the parent sample has the same sky distribution as the WISE eclipsing binary sample while the total number of sources (i.e. the sum of the weights) remain unchanged. The error bars in Fig.~\ref{fig:frac-Vt} are estimated using the Poisson statistics assuming no errors from the weights. The difference between the unweighted and weighted result is small, within $0.4$ of the error bars. 


The WISE and Gaia eclipsing binary samples are in excellent agreement in Fig.~\ref{fig:frac-Vt}: they show that the eclipsing binary fraction peaks at an intermediate tangential velocity ($\sim10^{1.5}$\kms), and decreases towards both low and high velocity end. With smaller error bars, the Gaia eclipsing binary sample constrains the peak to be in the bin of $10^{1.3-1.6}$\kms. As elaborated in more detail in Section~\ref{sec:population}, the difference of eclipsing binary fraction cannot be explained by the smaller sizes of stars with lower metallicities. This is the primary result of this paper: the fraction of short-period binaries is a strong function of kinematics.


We perform the Anderson-Darling test to quantify the significance of the difference in the distributions of tangential velocity between the sample of short-period eclipsing binaries and the comparison MS sample. The Kolmogorov-Smirnov test and the Anderson-Darling test give qualitatively the same result and here we quote the Anderson-Darling values because the Anderson-Darling test is more sensitive to tails of distributions. For WISE eclipsing binary sample, the p-value, the probability that two distributions are sampled from the same parent distribution, is 0.02. For Gaia eclipsing binary sample, the p-value is $6\times10^{-4}$. Therefore, the kinematic difference is statistically significant. 





In principle, different velocity components ($U$, $V$, and $W$) may provide different kinematic information for eclipsing binaries. For example, the velocity component in the direction of galactic rotation ($V$ component) would lag behind the disk as a result of asymmetric drift \citep{Dehnen1998, Reid2009}. However, since only tangential velocities are available for our sample, we find that decomposing the tangential velocity into $U$, $V$, and $W$ component suffers strongly from the projection and does not provide statistically meaningful constraints. Therefore, we focus on the results of tangential velocities in this paper.


\section{Potential systematics}
\label{sec:systematics}

\begin{figure}
	\centering
	\includegraphics[width=1.\linewidth]{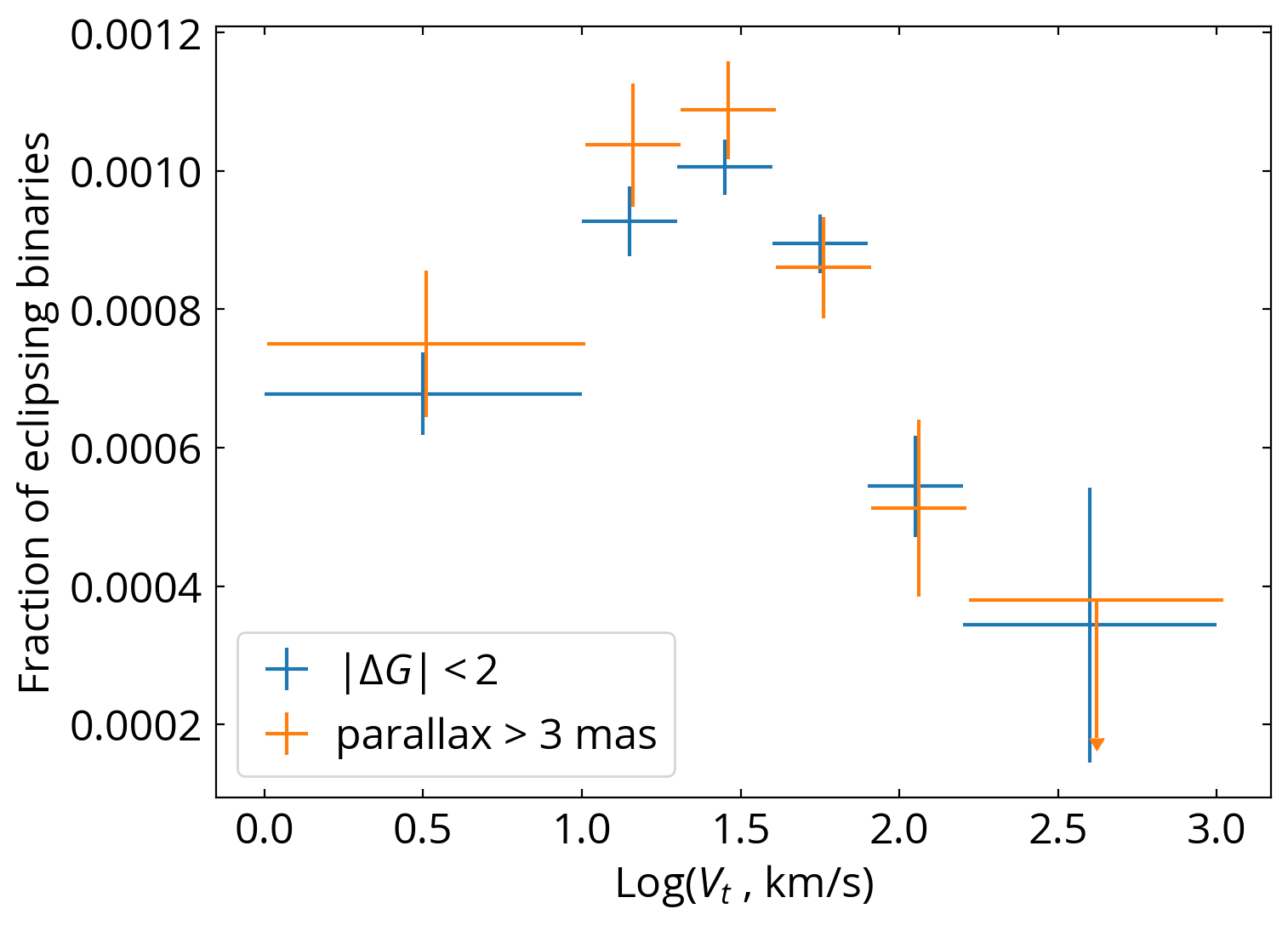}
	\includegraphics[width=1.\linewidth]{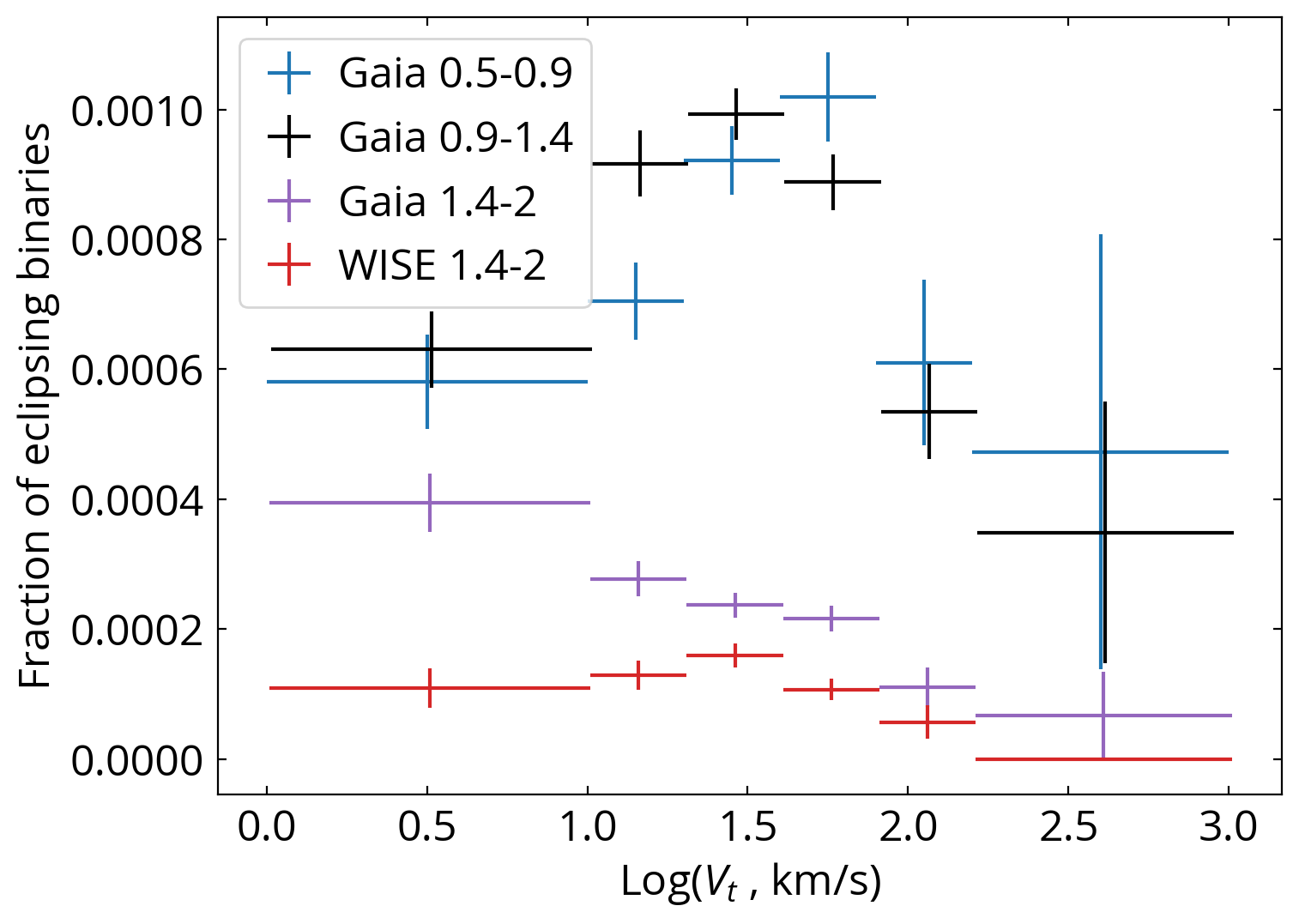}
	\caption{The same as Fig.~\ref{fig:frac-Vt}, but with different sample selections. Top panel: Different MS selection of $|\Delta G|<2$\,mag (blue crosses) and parallax $>3$\,mas (orange crosses), using the Gaia eclipsing binary sample. Bottom panel: Different color selections. BP$-$RP=0.5-0.9 (blue crosses) and BP$-$RP=0.9-1.4 (black crosses) from the Gaia eclipsing binary sample agree with each other very well. BP$-$RP=1.4-2 from the WISE eclipsing binary sample also shows a similar trend but with a lower eclipsing binary fraction compared to the bluer color ranges. BP$-$RP=1.4-2 from the Gaia eclipsing binary sample has a peak at the low velocity end, which is likely due to the active flaring from young late-type dwarfs.  
	}
	\label{fig:systematics}
\end{figure}

Because binaries are brighter than single stars, using magnitude cut could bias the sample. We use a volume-limited sample without any explicit magnitude cut, and in Fig.~\ref{fig:magpar-cut} we show that the binary fractions remain fairly flat over the entire range of parallax considered, meaning that eclipsing binaries within this distance range are well recovered. Furthermore, 95\% of our MS sample are brighter than 14.8\,mag in G-band while the limiting magnitude of Gaia DR2 is $\sim21$\,mag,  our criteria for the mean flux divided by its error do not imply any implicit magnitude cut.

The excellent agreement between the WISE sample and Gaia sample in Fig.~\ref{fig:frac-Vt} means that our results are not affected by the WISE cross-match nor the limit of period-finding algorithms. Furthermore, the difference in the observing strategies of WISE and Gaia (and the resulting differences in the sky distribution of binaries) does not appear to affect our result. The dependence of the binary fraction on $V_t$ cannot be explained by the covariance between velocity measurements and the variability. First of all, the tangential velocities are computed from proper motions and parallaxes with corrections from solar motions and the Galactic differential rotation, and there is no direct link to the photometry. If the observed dependence were due to the covariance between velocity and variability measurements, we would expect to see a monotonic relation in Fig.~\ref{fig:frac-Vt}, which is not the case.

Gaia DR2 uses the standard deviation of individual flux measurements to estimate the flux errors, so variable sources like eclipsing binaries may have lower {\tt phot\_g\_mean\_flux\_over\_error} (also depends on the number of observations). Therefore, if a stricter cut for the mean flux divided by its error is used, the eclipsing binary sample may be reduced. However, this only affects the completeness level but not the kinematics, so it is not expected to change the observational trend in Fig.~\ref{fig:frac-Vt}. To verify, we test a selection with the mean flux divided by its error only larger than 10 for G, BP and RP bands (instead of 50, 20, and 20), resulting in $\sim10$\% more eclipsing binaries but not affecting the conclusion in Fig.~\ref{fig:frac-Vt}.

The binaries in our sample only have separations of a few solar radii. For a solar-like contact binary at 100\,pc, the maximum angular separation of the binary is $\sim0.1$\,mas, and the observed angular separation is even smaller due to the orbital motion and the viewing angle. Therefore, the resulting astrometric noise is $\ll 0.1$\,mas, which is below Gaia's astrometric precision \citep{Lindegren2018}. If the short-period binary has a tertiary companion, the additional orbital motion of the short-period binary around the companion may affect the astrometric solution. The level of this effect depends on the separation (and the orbital period) and the mass ratio between the inner binary and the tertiary companion. If the tertiary is comparable to the binary in brightness, we would actually not see much motion since the center of brightness would be relatively still, but such objects will be rare because the tertiary would be diluting the eclipses and we would be much less likely to identify such sources. If the tertiary is faint and therefore we mainly measure the motion of the inner binary, then outer orbital period needs to be short enough ($\lesssim10$\,yrs) to contribute significant orbital velocity to the inner binary, but also long enough ($>2$\,yrs of Gaia DR2's observation) so that the motion of the inner binary can still be described by the single-star model used by Gaia DR2's astrometric solutions. Therefore, there is a very limited parameter space for tertiary contamination to have a significant effect on our results.



In Fig.~\ref{fig:systematics}, we establish the robustness of results to differences in sample selection. In the top panel, we use the Gaia eclipsing binary sample to test with a different MS selection of $|\Delta G| < 2$\,mag, and also with a closer sample of parallax $>3$\,mas (i.e. within 333\,pc). The results are nearly the same except that the fraction of binaries with parallax $>3$\,mas has larger error bars due to the smaller sample size. The results from the WISE eclipsing binary sample are similar so we do not repeat here.

In the bottom panel of Fig.~\ref{fig:systematics}, we test the fraction of eclipsing binaries with different color ranges, and therefore different mass ranges. We consider three BP$-$RP ranges: 0.5-0.9, 0.9-1.4 (the same in Fig.~\ref{fig:frac-Vt}), and 1.4-2. The first two agree with each other very well. Interestingly, the eclipsing binary fraction in the color range of 0.5-0.9 seems to peak at a higher velocity ($V_t=10^{1.6-1.9}$\kms). The WISE eclipsing binary sample with BP$-$RP=0.5-0.9 also shows similar results to the Gaia sample, so we do not repeat here. This blue sample may have some contamination from $\delta$ Scuti variables, especially that a large fraction (up to $\sim70$\%) of stars located in the $\delta$ Scuti instability strip are pulsating, but most of them are variable on levels of a few mmag, below our variability sensitivity \citep{Murphy2019}. Indeed, within 751 eclipsing binaries in this color range, only 4 are identified as high-amplitude ($>0.1$\,mag) $\delta$ Scuti/SX Phoenicis from Gaia's non-public light curves \citep{Rimoldini2019}, so such contamination is small and does not affect the results. 

The WISE eclipsing binary sample with BP$-$RP=1.4-2 (red crosses) shows a similar trend but with a lower eclipsing binary fraction compared to the bluer color ranges, which may be due to the combination of lower (eclipsing) binary fraction in low-mass stars \citep{Duchene2013}, the faintness of these stars, and their short periods below the classic Nyquist limit. The eclipsing binary fraction from the Gaia eclipsing binary sample with BP$-$RP=1.4-2 (purple crosses) peaks at the lowest velocity bin, with perhaps a slightly flattened trend at $V_t\sim 10^{1.6}$\kms. Because our Gaia eclipsing binary selection is based on the flux standard deviation but not the light curves, it is likely that this selection ends up with many actively flaring, young late-type stars. Due to the likely low completeness and high contamination of the reddest bin, we do not use it in our subsequent modeling.

\section{The lifetime of eclipsing binaries from the galactic model}
\label{sec:galactic-model}


The kinematics in Fig.~\ref{fig:frac-Vt} may be linked to the age of the stars. When stars form in the disk, they have similar circular velocity (with some offset, \citealt{Reid2009}) as the disk initially. As time goes by, stars are perturbed by structures like giant molecular clouds, transient spiral arms, bars, and flyby satellite galaxies, resulting in a higher velocity dispersion when stars age. The age-velocity dispersion relation has been widely studied in literature (e.g. \citealt{Nordstrom2004, Holmberg2009, Sharma2014a, Cheng2019}), and this relation is crucial for converting the kinematics into stellar ages.

Because the velocity dispersion monotonically increases with the stellar age, the average age of the stars in each tangential velocity bin in Fig.~\ref{fig:frac-Vt} is older with increasing velocities. Because the eclipsing binary fraction peaks at $10^{1.3-1.6}$\kms\ and drops at both lower and higher velocity ends, it means that the eclipsing binary fraction peaks at a certain stellar age, and is lower for younger and older populations. As a first-order approximation, we parameterize the eclipsing binary fraction as a function of stellar age using three parameters: intrinsic eclipsing binary fraction (IEBF), the time when the eclipsing binaries form ($t_0$), and the time when the eclipsing binaries disappear ($t_1$). $t_0$ and $t_1$ determine the overall trend of eclipsing binary fraction versus kinematics, and IEBF adjusts the normalization but does not affect the trend.

Fully modeling Fig.~\ref{fig:frac-Vt} requires a complete description from the Galactic model, including the Galactic star formation rate history, number densities and kinematics for different stellar populations. We use the Gaia DR2 mock catalog produced by \cite{Rybizki2018}. The Gaia DR2 mock catalog is generated using {\tt Galaxia} \citep{Sharma2011} that samples stars from a Besan\c con Galactic model \citep{Robin2003} with a realistic 3D dust extinction map \citep{Drimmel2003, Marshall2006, Green2015, Bovy2016b,Bovy2016}. Because we do not correct for dust extinction in our samples, they can be directly compared with the Gaia DR2 mock catalog, although dust extinction within 500 pc is not a strong effect (typically $A_{\rm V}<0.8$\,mag). The Gaia DR2 mock catalog also provides the ages and metallicities of the sampled stars, which is necessary for us to model the eclipsing binary lifetime. 

We select stars from the Gaia DR2 mock catalog using the same color and absolute magnitude criteria as our sample, i.e. $0.9<$BP$-$RP$<1.4$, $|\Delta \rm G |< 1.5$, and parallax $>2$\,mas. The Gaia DR2 mock catalog itself does not simulate the stellar binaries, so for sources that are supposed to be binaries, their luminosities are underestimated by $\le 0.75$\,mag. Our absolute magnitude selection of $|\Delta \rm G |< 1.5$ ensures that such systems are selected in both our eclipsing binary samples from observations and from the mock catalog. We assign weights to the stars in the mock catalog so that their sky distribution is the sample as our observational Gaia EB sample. The tangential velocities are corrected by removing the solar motion and the Galactic differential rotation.

\begin{figure}
	\centering
	\includegraphics[width=.9\linewidth]{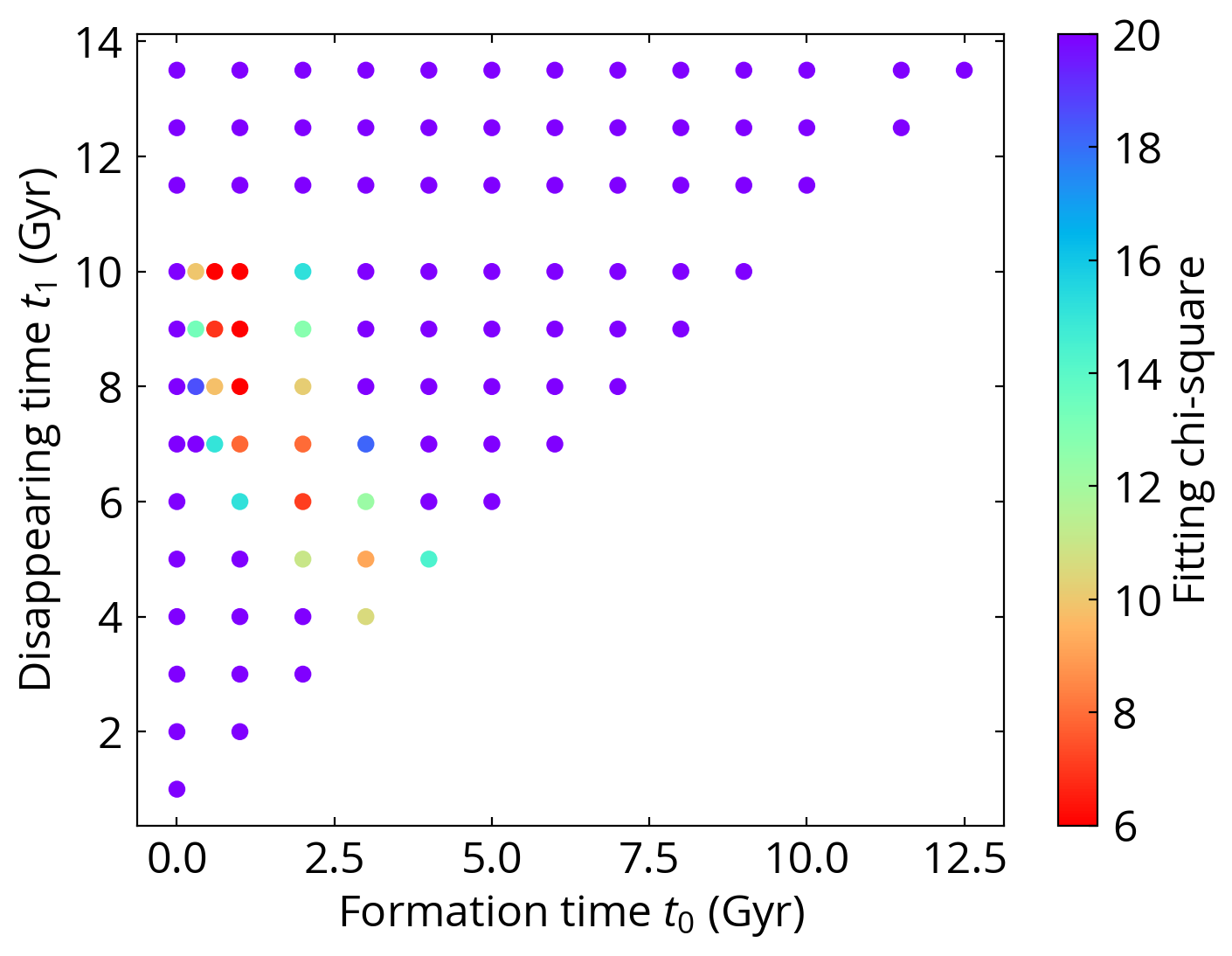}
	\caption{The model grids for the formation time ($t_0$) and disappearing time ($t_1$) of eclipsing binaries, color-coded by the chi-square of the best fit. The result constrains the formation time to be $t_0=$0.6-3\,Gyr and the disappearing time $t_1=$5-10\,Gyr, with accepted models roughly following the relation $t_0 + 0.4 t_1 \sim5$\,Gyr. }
	\label{fig:model-grid}
\end{figure}

We sample a grid of formation time ($t_0$) and disappearing time ($t_1$) shown in Fig.~\ref{fig:model-grid}. For each combination of $t_0$ and $t_1$, we feed them into the Gaia DR2 mock catalog, and using the stellar ages recorded in the mock catalog, we compute the preliminary (preliminary because it has not considered the IEBF) eclipsing binary fractions weighted by the sky distribution as a function of tangential velocity. Then the preliminary eclipsing binary fractions are fit to the observed WISE-selected EB sample to determine the best-fit IEBF and the corresponding linear chi-squared costs, presented by the color coding in Fig.~\ref{fig:model-grid}.

\begin{figure*}
\centering
\includegraphics[width=.45\linewidth]{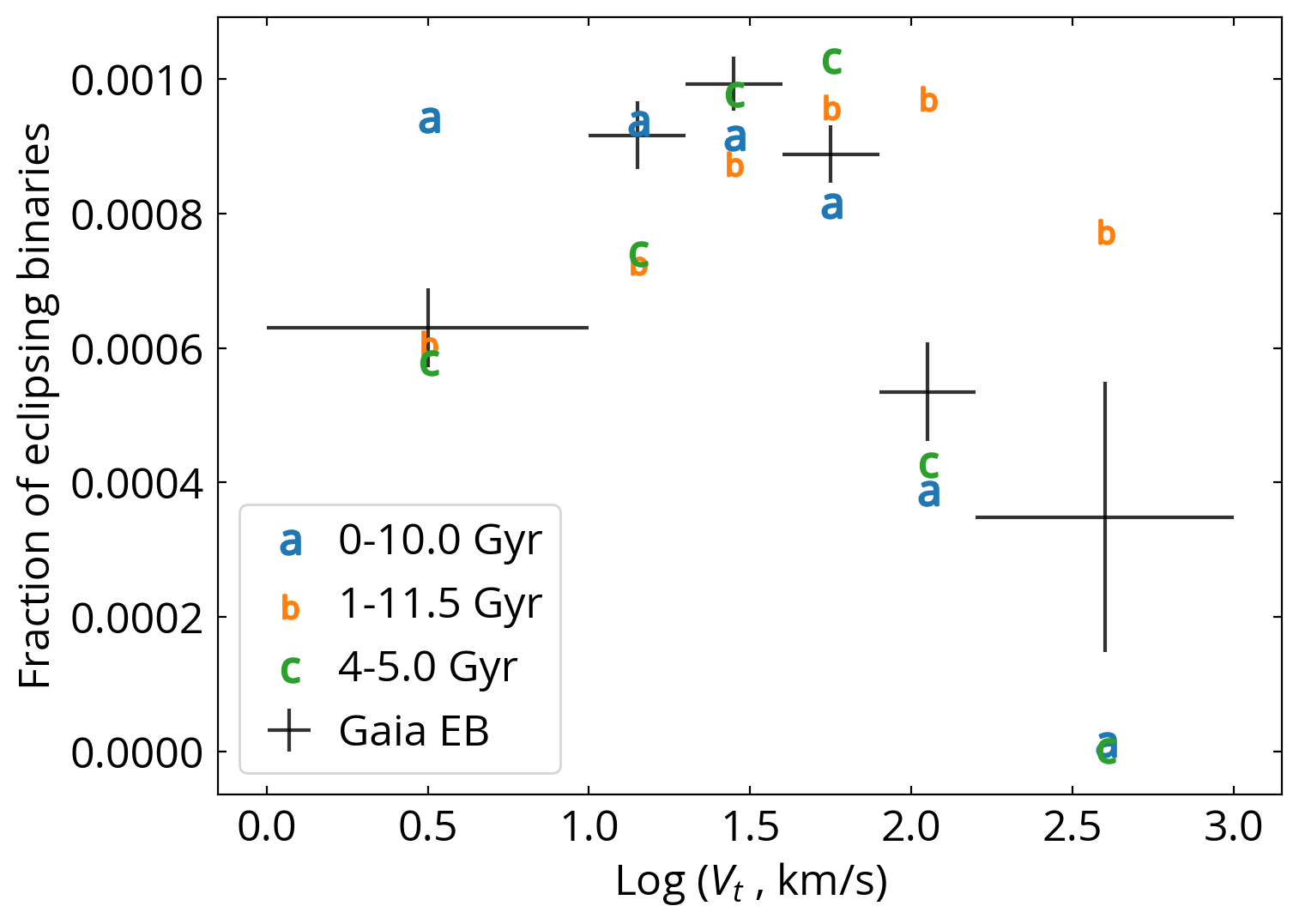}
\includegraphics[width=.45\linewidth]{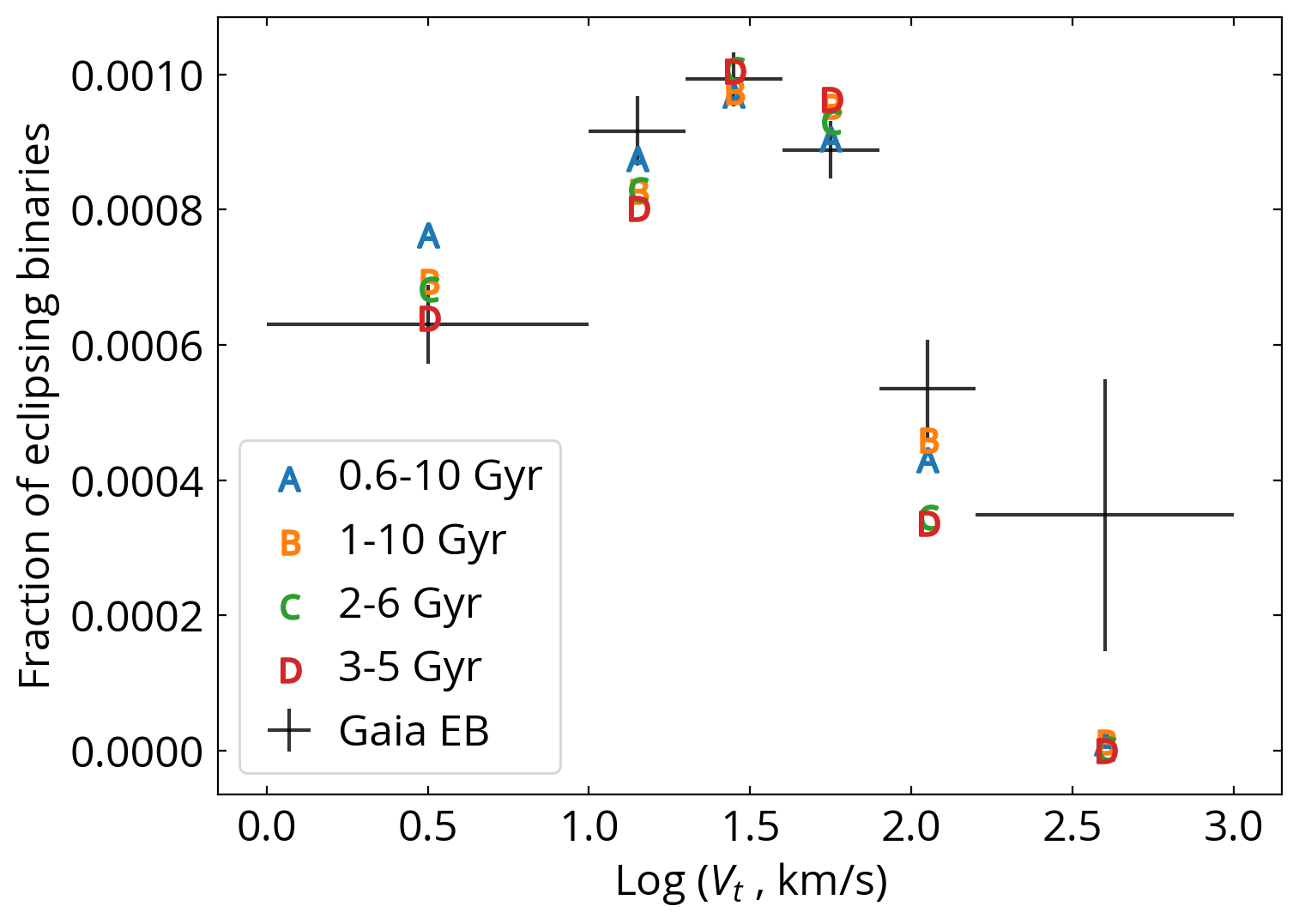}
\caption{Examples of rejected models (left panel) and accepted models (right panel) for the lifetime of eclipsing binaries. }
\label{fig:model-accept-reject}
\end{figure*}

Fig.~\ref{fig:model-grid} shows that models with $t_0=0$\,Gyr and those with $t_1\ge12$\,Gyr can be rejected. We avoid using fits with 11\,Gyr and 13\,Gyr because these are the ages of thick-disk stars and halo stars in the mock catalog. We present some rejected examples in the left panel of Fig.~\ref{fig:model-accept-reject}. The observed drop of eclipsing binary fractions on the low-velocity end leads to rejection of models with $t_0=0$\,Gyr because such models can naturally only produce monotonically decreasing eclipsing binary fraction with increasing velocity (since the mean stellar ages monotonically increase with increasing velocity). On the other end of the distribution, models with $t_1\ge12$\,Gyr (i.e. when binaries can only disappear at an age above that of thick disk) make the eclipsing binary fraction too high in the velocity bins $>100$\kms, for example the model (b) in the left panel of Fig.~\ref{fig:model-accept-reject}.

Fig.~\ref{fig:model-grid} presents the accepted models where $t_0\sim0.6$-3\,Gyr and $t_1=$5-10\,Gyr, and the accepted $t_0$ and $t_1$ roughly follow a relation of $t_0 + 0.4 t_1 \sim 5$\,Gyr. Some examples of the accepted models are shown in the right panel of Fig.~\ref{fig:model-accept-reject}. They all successfully reproduce the overall trend of eclipsing binary fractions as a function of velocity.



The main uncertainty in these models lies in the Galactic descriptions used, including the star formation history, the adopted age-velocity dispersion relation, kinematics descriptions for different stellar populations (thin disk, thick disk, and halo), etc. These models are currently calibrated by the entirety of data from Galactic surveys. The number of free parameters involved is too large for us to investigate the uncertainty if a different Galactic description is used. Another uncertainty is the step-function-like lifetime model. While it is a reasonable first step, it is likely too simplistic. Because the uncertainties are mostly due to the model assumptions rather than due to measurement uncertainties, we do not pursue a best fit nor the Markov chain Monte Carlo procedure. Even though the modeling uncertainties are still unclear, the observed relation between eclipsing binary fraction and velocity can be successfully reproduced using the state-of-art Galactic descriptions. 


\section{Discussion}
\label{sec:discussion}

\subsection{Different properties between thin-disk, thick-disk and halo stars?}
\label{sec:population}

First, we attempt to determine if our results in Fig.~\ref{fig:frac-Vt} can be explained by the different eclipsing binary fractions in the thin-disk, thick-disk, and halo stars, without explicit consideration of stellar ages. Qualitatively, it is difficult because for $V_t<100$\kms, the sample is dominated by thin-disk stars and therefore a constant eclipsing binary fraction in thin-disk stars cannot explain the trend at $V_t<100$\kms\ in Fig.~\ref{fig:frac-Vt}. For $V_t>100$\kms, the thick-disk and halo stars start to dominate the sample so the decreasing eclipsing binary fraction might be linked to the different eclipsing binary fractions in different stellar populations.

\begin{figure*}
	\centering
	\includegraphics[height=.34\linewidth]{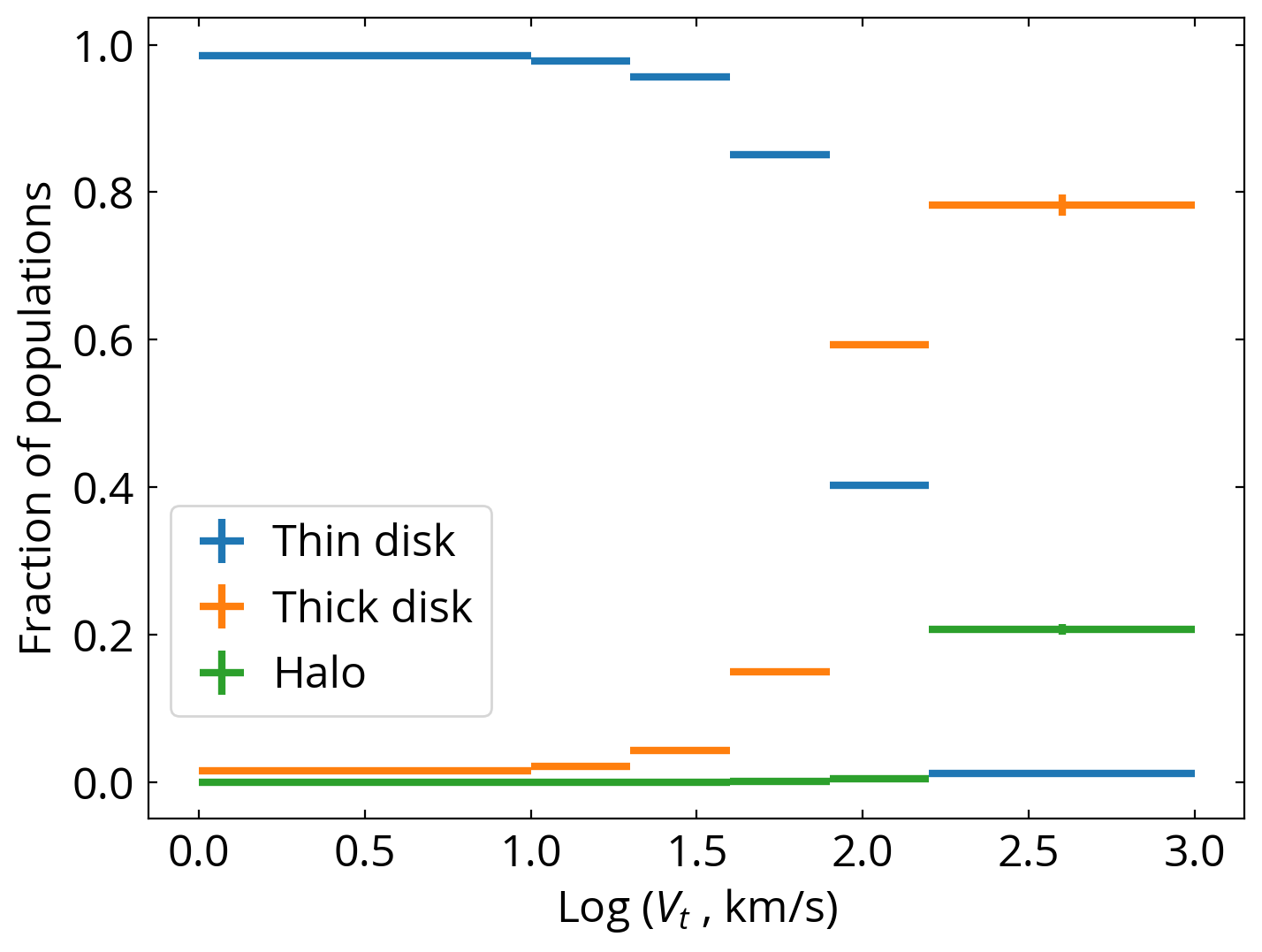}
	\includegraphics[height=.34\linewidth]{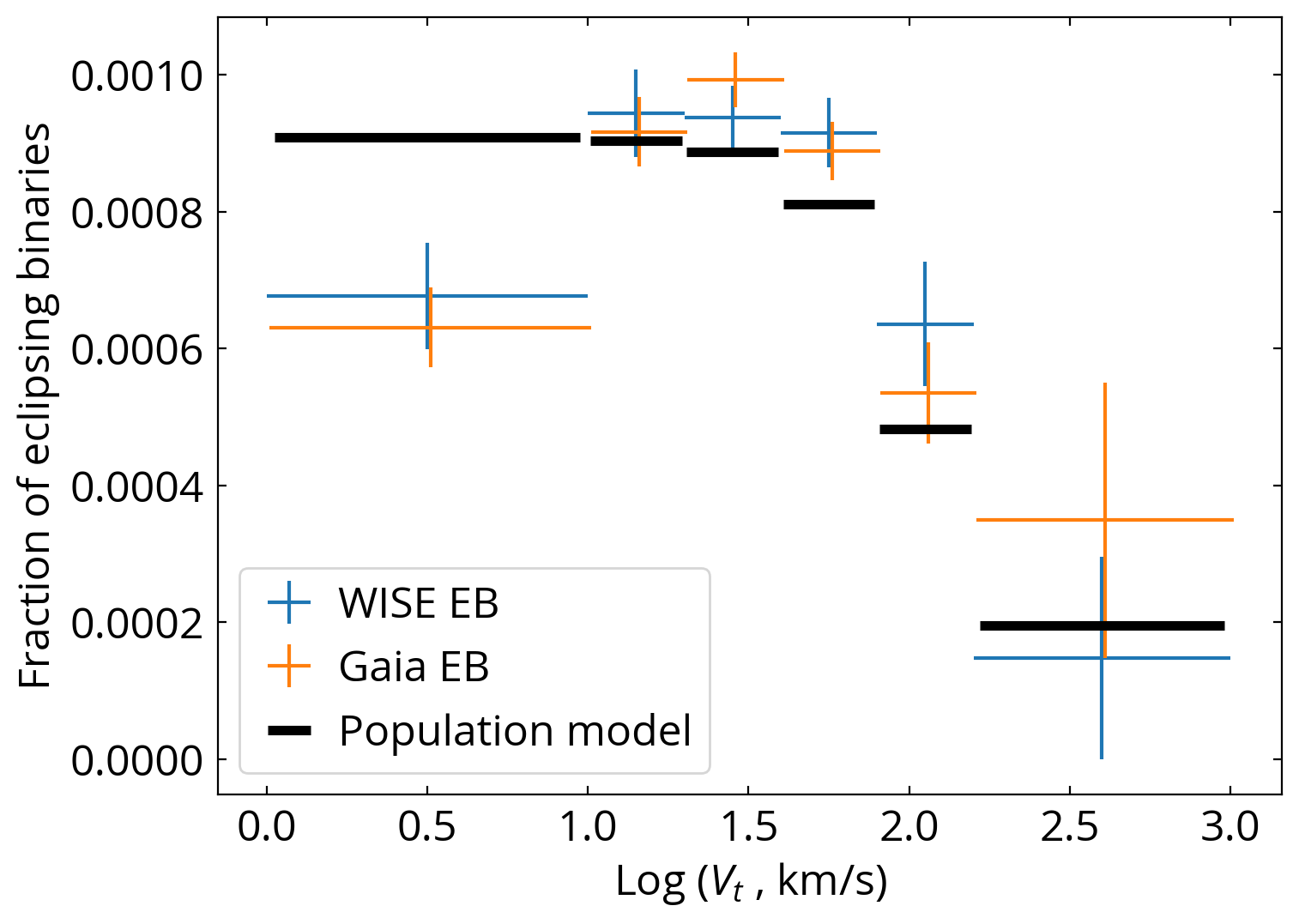}
	\caption{Left: fraction of the population (thin-disk, thick-disk, and halo stars) in each tangential velocity bin from the Galactic model. Right: eclipsing binary fraction versus tangential velocity, with a best-fit model (black horizontal bars) that considers different eclipsing binary fractions in each population. Age is not explicitly taken into account in the model. The best fit gives that the eclipsing binary fraction is $\sim10$ times smaller in thick-disk (and probably halo) stars than in thin-disk stars. The population model can reproduce the observational trend on the high velocity end, but not on the low velocity end. }
	\label{fig:populations}
\end{figure*}

The left panel of Fig.~\ref{fig:populations} presents the fractions of each stellar populations in each tangential velocity bins with the same selection in the H-R diagram as Fig.~\ref{fig:HR}, weighted by the sky distribution of the Gaia eclipsing binary sample. The fractions of each stellar populations are derived from the Gaia Mock DR2 Catalog. The fractions of stellar populations in a tangential velocity bin can also be derived by considering the location distribution in the H-R diagram because thin-disk, thick-disk, and halo stars are located differently in the H-R diagram due to the difference in metallicity (e.g. Fig. 21 and 22 in \citealt{Gaia2018Babusiaux}). We use this method to obtain the fractions of each stellar populations from the Gaia data, with a similar result to that from the Gaia Mock DR2 Catalog. Fig.~\ref{fig:populations} shows that $>90$\% of the sample are thin-disk stars for $\log(V_t)<10^{1.7}$\kms, and $>60$\% are thick-disk stars for $\log(V_t)>10^{1.9}$\kms. Halo stars become the dominant population ($>50$\%)  when $\log(V_t)>10^{2.3}$\kms, but the fraction of halo stars is reduced to $19$\% for $\log(V_t)>10^{2.2}$\kms.

The right panel of Fig.~\ref{fig:populations} shows the best-fit model that considers different eclipsing binary fractions for thin-disk, thick-disk, and halo stars. Because the halo stars only compose 19\% of the sample in the highest velocity bin, its eclipsing binary fraction is not well constrained and hence we assume that the thick-disk stars and halo stars have the same eclipsing binary fractions during the fitting. The best-fit eclipsing binary fraction is $0.111\pm0.003$\% for thin-disk stars, and $0.012\pm0.007$\% for thick-disk and halo stars. Therefore, without the consideration of ages, the eclipsing binary fraction of thin-disk stars is $\sim10$ times larger than the one of thick-disk and halo stars. 

The best-fit model in the right panel of Fig.~\ref{fig:populations} is not able to reproduce the rising eclipsing binary fraction at $\log(V_t)<10^{1.5}$\kms. It is expected because thin-disk stars dominate in this velocity range and the model just reflects the eclipsing binary fraction of the thin-disk stars. Therefore, the eclipsing binary fraction of thin-disk stars cannot simply be a constant as a function of age.

The difference in eclipsing binary fractions between thin-disk stars and thick disk stars (and possibly halo stars) can be due to several factors. Because thick-disk and halo stars are older than thin-disk stars, the different eclipsing binary fraction may be the consequence of the eclipsing binary lifetime like Fig.~\ref{fig:model-accept-reject}. Thick-disk and halo stars are more metal-poor compared to thin-disk stars, and the effect of metallicity is discussed in the next section. Halo stars may be accreted from infalling satellite galaxies instead of forming in the Milky Way, and therefore their formation environment can be different. The different eclipsing binary fractions might also result from the difference in physical properties between populations. For example, at fixed colors, metal-poor stars are smaller in size than metal-rich stars. Because the probability of being an eclipsing system is proportional to $R/a$, where $R$ is the size of the star and $a$ is the semi-major axis of the binary, smaller sizes of thick-disk stars might reduce the eclipsing binary fraction. However, we consider it unlikely. At the color of our sample, thick-disk stars are $\sim0.3$\,mag fainter than thin-disk stars, or a factor of $\sim0.87$ smaller in the stellar radius. To reduce the eclipsing binary fraction by a factor of 10, thick-disk stars need to have a separation distribution 9 times wider than thin-disk stars. It is unlikely because that would make the period distribution of thick-disk stars $\sim$30 longer than thin-disk stars. 

\begin{figure}
	\centering
	\includegraphics[width=.9\linewidth]{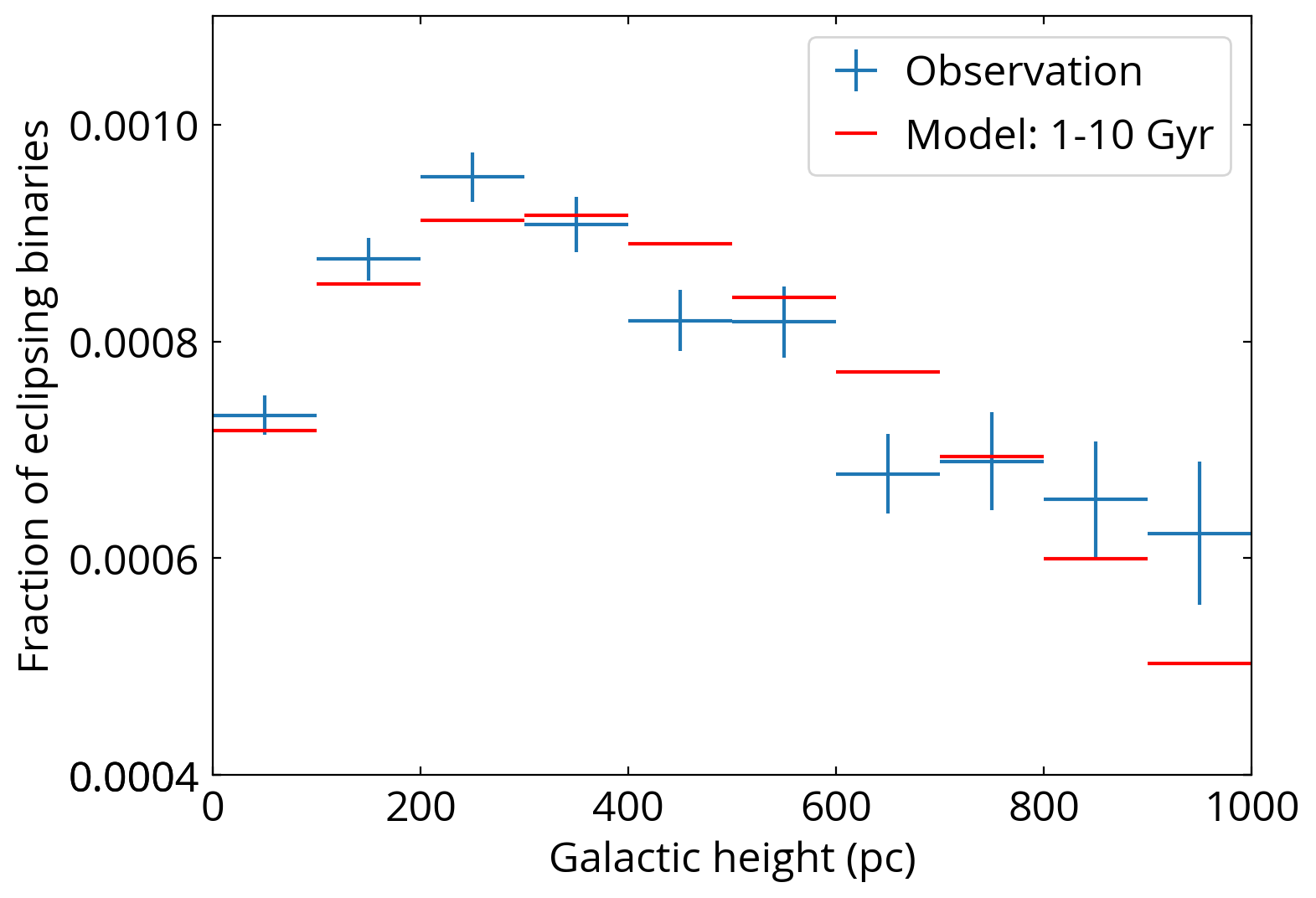}
	\caption{The eclipsing binary fraction as a function of the Galactic height. The blue crosses are the observational result where the horizontal segments indicate the bin sizes and the vertical segments are the errors of eclipsing binary fractions. The red horizontal bars are the model where the length of the bars is the bin size. The model uses a contact binary lifetime of 1 to 10 Gyr. The increasing eclipsing binary fraction with respect to Galactic latitudes at Galactic latitudes $<300$\,pc is due to the delayed formation of short-period binaries, and the deceasing eclipsing binary fraction at Galactic latitudes $>300$\,pc shows that the thick disk has a much lower eclipsing binary fraction.}
	\label{fig:gal-height}
\end{figure}

The difference in the eclipsing binary fractions between the thin and thick disk is best demonstrated by the dependence of binary fraction on the Galactic height in Fig.~\ref{fig:gal-height}. We slightly modify the sample selection here so that the sample remains complete to distances of $\sim1$\,kpc with sufficient statistical sample sizes. In this plot, the main-sequence sample is selected by $0.5<$BP$-$RP$<1.1$, $|\Delta \rm G |< 1.5$, and parallax $>0.8$\,mas, and the eclipsing binaries are selected using $\log(f_G^2) > -2$. The Galactic height is computed by $d\times |\sin(b)|$, where $d$ (inverse of parallaxes) is the distance of the star from the Sun and $b$ is the Galactic latitude. Since we are interested in the change of eclipsing binary fractions on scales of $>100$\,pc, we do not correct for the height of the Sun above the Galactic plane, which is $\sim15$ pc \citep{Binney1997, Widmark2019}. We also show comparison with a $t_0=1$\,Gyr and $t_1=10$\,Gyr model, where we have modified the model in agreement with the selection used in Fig.~\ref{fig:gal-height}.

Fig.~\ref{fig:gal-height} demonstrates that the eclipsing binary fraction decreases when the Galactic height $\gtrsim300$\,pc, where the thick-disk stars become increasingly dominant. This strengthens our conclusion that the thick disk has much lower eclipsing binary fraction than the thin disk. The increasing eclipsing binary fraction with increasing Galactic height at Galactic heights $< 300$\,pc is the consequence of the delayed formation of these short-period binaries. These trends are in excellent agreement with the model constructed independently based on the kinematic information. The strong dependence of short-period binary fraction on age, and consequently their kinematics and their Galactic height, explains the dependence on Galactic latitudes of eclipsing binary fraction seen in the literature (e.g. \citealt{Prsa2011,Slawson2011,Kirk2016}).

\cite{Latham2002} find that there is no significant difference in the period distribution of spectroscopic binaries between disk stars and halo stars. Out of 156 objects with robust orbital solutions in their sample, the shortest period is 1.93\,day, and only 7 (4.5\%) have periods $<10$\,days. Therefore, it is likely that our results are different from theirs because we are probing a much shorter period population ($<$1\,day) in which stronger evolutionary effects may be expected.

To summarize, while the declining eclipsing binary fraction at $\log(V_t)>10^{1.5}$\kms\ suggests a much smaller eclipsing binary fraction in thick-disk and possibly halo stars, the rising eclipsing binary fraction at $\log(V_t)<10^{1.5}$\kms\ is best explained by a delay in formation of eclipsing binaries compared to the formation of their components. We discuss possible causes for the delayed formation time and for the disappearing time in the following sections.


\subsection{Metallicity}
\label{sec:metallicity}

Recent studies have shown that the close-binary fraction (periods $<10^4$\,days; separation $<10$\,AU) increases with decreasing stellar metallicity \citep{Grether2007,Yuan2014,Badenes2018,Moe2019}, consistent with formation of cose binaries due to disk fragmentation \citep{Tanaka2014}. While our eclipsing binary sample has periods much shorter than their close binaries, we investigate if our results can be explained by the metallicity dependence.

\begin{figure}
	\centering
	\includegraphics[width=.9\linewidth]{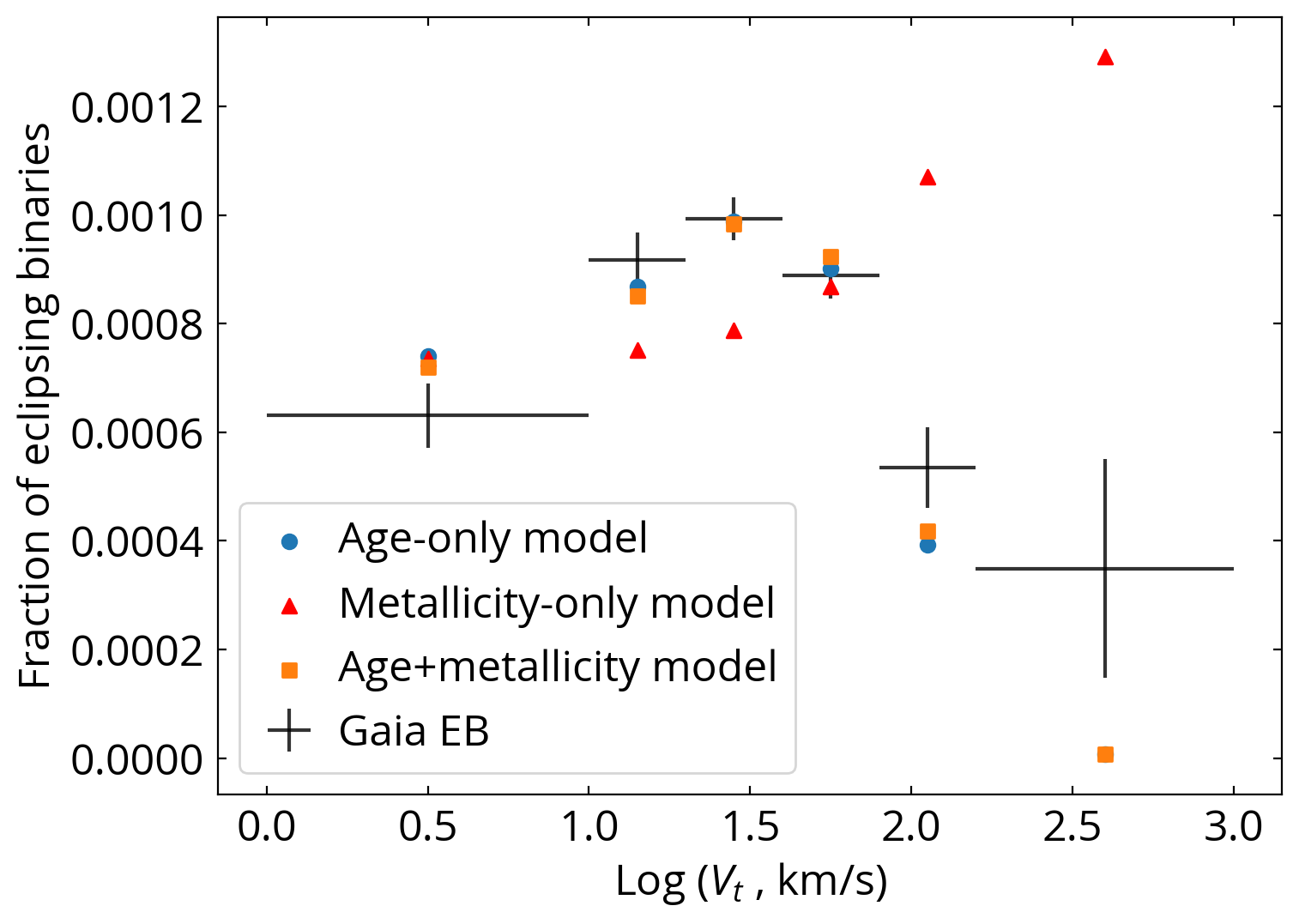}
	\caption{The eclipsing binary fraction versus tangential velocity, with models that take metallicity into account. The age-only model uses $t_0=1$\,Gyr and $t_1=8$\,Gyr. The metallicity-only model (red triangles) cannot reproduce the observed trend. }
	\label{fig:model-metallicity}
\end{figure}

Fig.~\ref{fig:model-metallicity} presents the models that take metallicity into account. We adopt the close binary fraction as a function of stellar metallicity from \cite{Moe2019}, and determine the normalization during the fitting (because not all close binaries are eclipsing binaries). The red triangles in Fig.~\ref{fig:model-metallicity} show the model that only includes metallicity effect but not age. The resulting binary fraction is inconsistent with the observation in two ways. First, for radial velocity $V_t<10^{1.5}$\kms, the metallicity-only model cannot reproduce the rising binary fraction as steep as the observation. This is because there is no strong metallicity-age relation for stellar ages $\lesssim5$\,Gyr \citep{Casagrande2011,Bensby2014,SilvaAguirre2018}. Second, the metallicity-only model does not have the decreasing binary fraction at velocity $V_t>10^{1.6}$\kms. Therefore, taking at face value the metallicity dependence by \cite{Moe2019}, our results cannot be explained by metallicity alone. 

Fig.~\ref{fig:model-metallicity} also presents a model which includes both metallicity and age (orange squares). The adopted lifetime parameters are $t_0=1$\,Gyr and $t_1=8$\,Gyr. The metallicity+age model shows a slight improvement in the velocity bin at $V_t=100$\kms over the age-only model. Since this velocity bin is dominated by thick-disk stars, the goodness of the fit relies on the model descriptions, and therefore we do not favor the metallicity$+$age model because of its slight improvement.

We conclude that metallicity dependence is not able to explain the observational trends in eclipsing binary fraction versus kinematics. It is probably due to our sample focusing on the shortest period end, and mechanisms of orbital migration may make this sample more sensitive to the stellar ages. Binaries with longer periods (e.g. spectroscopic binaries) may not experience all the mechanisms of orbital migration, and therefore the effect of age is not prominent. 



\subsection{The formation of eclipsing binaries}
\label{sec:formation}

Our results show that short-period eclipsing binaries form with a delay of $\gtrsim 0.6$\,Gyr. Because the size of pre-MS stars is much larger than zero-age MS stars, the separation between two stars in a binary must be larger than these eclipsing binaries in the beginning. Therefore, the formation delay is due to the orbital migration that a binary undergoes to lose the orbital angular momentum until the orbital period is $\lesssim1$\,day. 

Binaries can lose their orbital angular momentum through the energy dissipation in the pre-MS phase \citep{Bate1998,Tohline2002,Moe2018}, through the angular momentum exchange with a distant tertiary and tidal effects \citep{Kiseleva1998, Fabrycky2007}, and magnetic braking \citep{Stepien1995}. These mechanisms dominate different stages of orbital migration over different timescales.

Our estimated formation time of short-period binaries places strong constraints on the binary evolution theory. Because the pre-MS phase happens on a very short timescale ($\lesssim$ a few Myr), it does not fulfill the delayed formation time of $\sim1$\,Gyr. This means that short-period binaries cannot be produced only by the interaction in the pre-MS phase. The timescale of the Kozai-Lidov effect to produce short-period binaries depends on the initial conditions of the binary, including the initial separations and initial eccentricity of inner binary, ranging from $\sim$\,Myr to a few hundred Myr \citep{Fabrycky2007,Perets2009}. Since the orbits of binaries are circularized at orbital periods of $\sim10$\,days \citep{Latham2002, El-Badry2018}, it is difficult for the Kozai-Lidov mechanism to further reduce the orbital periods after that, and therefore other processes such as magnetic braking may be needed to complete the last step of orbital migration. Magnetic winds can bring detached binaries from periods of 5 days to contact binaries over a few Gyr \citep{Stepien1995,Yakut2008}. This timescale seems to agree with our constraint, but our upper limit of $\sim3$\,Gyr for the formation time places a strong constraint on the possible parameter space. 

While the delayed formation of short-period binaries is mostly determined by the magnetic braking, it does not mean that magnetic braking is the only process during the orbital evolution. In particular, because effective magnetic braking requires a small initial separation ($\lesssim 5$\, day, \citealt{Stepien1995}), pre-MS interaction and the Kozai-Lidov effect may still play an important role to bring binaries to orbital periods within $\sim5$\,days at an earlier stage.

Fig.~\ref{sec:systematics} shows that the bluer color selection of BP$-$RP=0.4-0.9 has an eclipsing binary fraction peaking at a higher tangential velocity than the redder sample, indicating a potential mass dependence. While it requires a more detailed analysis, such mass dependence may be an important clue on the dominant orbital migration process. For example, magnetic winds require the presence of subphotospheric convection zones that are only in low-mass stars ($\lesssim 1.3$\Msun). Therefore, if magnetic winds are the main cause for the delayed formation time in the color range of BP$-$RP=0.9-1.4, we may expect a longer delayed formation time for high-mass short-period binaries. The mass dependence of fragmentation during the proto-stellar phase may also play an important role (e.g. \citealt{Kratter2006}).

Our constraint of the formation time $\gtrsim0.6$\,Gyr is consistent with observations that no short-period binaries ($P<1$\,day) are found in T Tauri stars and young clusters \citep{Mathieu1994,Melo2001,Hebb2010}. While short-period eclipsing binaries are easy to detect if they exist, none is found with periods $<1$\,day in Hyades and Pleiades \citep{Torres2003, David2015, David2016}, and only one is found in Praesepe \citep{Rucinski1998, Zhang2009}. Further investigation is required to determine the true eclipsing binary fraction in open clusters for comparison with our results. Historically, RS Cha, with an orbital period of 1.67 days, was the shortest-period pre-MS eclipsing binary \citep{Alecian2007}, but Kepler K2 observations of Upper Scorpius (age of 5-7 Myr) have recently revealed three shorter period pre-MS eclipsing binaries EPIC 204506777, 203476597, and 202963882 where the orbital periods are 1.63, 1.44, and 0.63 days, respectively \citep{David2019}. The latter system is composed of two low-mass stars with 0.29\Msun and 0.20\Msun, much lighter than our sample.



\subsection{The disappearance of eclipsing binaries}

In Sec.~\ref{sec:population}, we show that thick-disk and halo stars dominate the sample for $\log(V_t)>10^{1.9}$\kms, and a factor of $\sim10$ smaller eclipsing binary fraction in thick-disk and halo stars can explain the observed declining eclipsing binary fraction at the high-velocity end. One possibility is that the eclipsing binary lifetime is shorter than the MS lifetime of these thick-disk and halo stars, making the eclipsing binary fraction in these populations much smaller compared to thin-disk stars. In this scenario, our results suggest that the disappearing time is between 5-10\,Gyr, depending on the formation time. Although the disappearing time is not well constrained, we discuss some possible scenarios that limit the lifetime of eclipsing binaries.


Contact binaries may end up as stellar mergers. \cite{Tylenda2011} report a stellar merger of a contact binary V1309 Scorpii, although its progenitor is probably at the beginning of the red giant branch and not a MS considered here. The merging product may eventually become a blue straggler \citep{Robertson1977}. By using binary evolutionary models, \cite{Stepien2015} show that some contact binaries can merge and become blue stragglers within the age of globular clusters ($\le 13$\,Gyr), and they suggest that this formation track may constitute a substantial fraction of all blue stragglers in globular clusters.

Our sample of BP$-$RP=0.9-1.4 has a MS lifetime longer than 14\,Gyr. If the declining eclipsing binary at $\log(V_t)>10^{1.9}$\kms\ is due to the stellar mergers of contact binaries, our results imply that the majority of short-period MS binaries are destroyed before the age of the thick disk ($\sim11$\,Gyr) and before the end of their own MS lifetime. This scenario can be tested by searching for high-velocity merging products, for example field blue stragglers.





A few other possibilities may reduce the binary fraction in old stars. One possibility is that their lower metallicity makes the orbital migration more inefficient, for example by suppressing the formation of triples, but this interpretation is disfavored by \cite{Moe2019} where they show that the triple star fraction increases with decreasing metallicity. Alternatively, these old stars were originally in binaries with more massive stars, which have evolved into compact objects (white dwarfs, neutron stars, or black holes) and therefore only the originally less-massive stars are visible now. It is not impossible because O- and B- binaries with periods $<20$\,days seem to favor modest mass ratios ( $q\sim0.5$; \citealt{Moe2017}). If some of the high-velocity stars in our sample indeed have invisible companions with periods $<20$\,days, the radial velocity variation is on orders of $\sim10$\kms, which is detectable by Gaia's radial velocity measurements.





\subsection{Interpretation of the formation time and disappearing time}

Our results are consistent with the age estimate of contact binaries in literature. Kinematic studies show that the age of contact binaries is of several Gyr \citep{Guinan1988, Bilir2005}. \cite{Yildiz2014} estimate the age of $\sim4.5$\,Gyr for W UMa binaries based on the stellar model \citep{Yildiz2013a} and kinematics. These estimates are consistent with our formation time and disappearing time. 


One distinction between this work and the literature is that we constrain the formation time and the disappearing time, not just the average age of eclipsing binaries. We emphasize that the formation time and disappearing time of short-period binaries are constrained in an average sense because of the use of the simple lifetime model. Our results do not imply that all eclipsing binaries form and disappear at the same time. In fact, it is very likely that the formation time itself is a wide distribution because the orbital migration processes, especially the Kozai-Lidov mechanism, is sensitive to the initial conditions (e.g. \citealt{Perets2009}).


Because the formation time and disappearing time are derived in an average sense, their difference ($t_1-t_0$) may not directly reflect the lifetime of the contact phase. Such timescale of the contact phase is rather uncertain, with some estimates ranging from $0.1$\,Gyr \citep{vantVeer1979, Eggen1989} to $\sim10$\,Gyr \citep{Mochnacki1981}. If the contact phase is short ($<1$\,Gyr), then $t_1-t_0$ is mainly related to the distribution of the formation time. If the contact phase can last for $>$ a few Gyr, $t_1-t_0$ may be able to constrain the timescale of the contact phase.



\subsection{Alternative explanations?}

In this section, we explore other possibilities that might explain the result. First, our results cannot be explained by different binary properties in different mass of young clusters because typical 1-D velocity dispersion in young clusters is $\lesssim 10$\kms\ (e.g. \citealt{Larsen2004}) while our result shows that eclipsing binary fraction peaks at $\sim30$\kms.

When binaries evolve to contact binaries, their surface temperatures may change, and the mass transfer may alter their eclipse depths and light curve profiles \citep{Yakut2005, Stepien2012}, moving them into or out of our sample selection. The color evolution from detached to contact binaries is a long-standing open question, and in some models the effective temperature of contact binaries oscillates without reaching an equilibrium (the thermal relaxation oscillation model; \citealt{Lucy1968,Flannery1976, Webbink1976,Yakut2005}).

While it is difficult to quantify these effects at the moment, we argue that our results are not strongly affected by such color evolution. (1) Among close binaries, there is an excess of nearly equal-mass binaries, so called twins \citep{Tokovinin2000,Pinsonneault2006,Raghavan2010}. Among wider binaries with primary masses 0.8-1.2\Msun\ and $P=10^{0.5-1.5}$ day, the fraction of nearly equal ($q>0.95$) binaries is $\sim38$\% \citep{Moe2017}, and it is likely to be even larger at shorter periods. When these nearly equal-mass binaries evolve to contact binaries, they do not experience significant color evolution. (2) Consider an effective observed temperature $T_{\rm eff}$ of an unresolved, detached binary defined as $R_1^2T_1^4 + R_2^2T_2^4 = (R_1^2+R_2^2)T_{\rm eff}^4$, where $R_1$ and $R_2$ are the radii and $T_1$ and $T_2$ are the temperatures of the two component stars, respectively. With the scaling relation of $R\propto M^{0.8}$ and $T\propto M^{0.54}$ for low-mass stars and $M$ is the mass of a star \citep{Lamers2017}, the minimum observed temperature happens at $T_{\rm eff}=0.8 T_1$ when $q=0.6$.  For a smaller mass secondary star ($q<0.6$), $T_{\rm eff}$ increases toward $T_1$ because the light coming from the secondary becomes negligible. During the contact phase, one extreme case is that two stars reach the same temperature $T_{\rm contact} = T_1$ from $T_{\rm eff}$, then it results a maximum color change of $\Delta$BP$-$RP$\sim0.5$ (from the simulated colors in PARSEC isochrones). In reality, $T_{\rm contact} < T_1$ and the color change is smaller ($\Delta$BP$-$RP$<0.5$). If our kinematic result is due to the color evolution, it means that the result would depend on how wide our color selection is. We test our result with a color selection of BP$-$RP$=$0.5-1.5 and the result remains the same, and therefore our result is not the consequence of such color evolution.

Another selection effect is unresolved tertiary companions. If the unresolved tertiary contributes non-negligible fluxes, it may change the apparent color of the binary and/or reduce the photometric variability, which would affect our selection. However, if the effect of unresolved tertiary companions is not a strong function of age (and therefore kinematics), then it only reduces the completeness of our sample without biasing the results. 

There is one interesting scenario where unresolved tertiary companions have an age dependence, in the sense that there is an excess of short-lived, unresolved tertiary companions around the short-period main-sequence binaries. In this case, we would only see the short-period binary after the short-lived, unresolved tertiary companion dies, which would explain the increasing eclipsing binary fraction at the low-velocity end in Fig.~\ref{fig:frac-Vt}. From our model, the lifetime of these short-lived tertiary companions needs to be shorter than 3\,Gyr, corresponding to stellar masses $>1.5$\Msun. This scenario contradicts previous studies which show that most of the tertiary companions are lighter than the total mass of the inner binaries \citep{Tokovinin2006,Borkovits2016}. This scenario would also imply that nearly all short-period binaries have close white dwarf companions, which has not been reported. 

Another scenario is that the tertiary companion is at a short separation from the inner binary at the young age and is blended with it, but becomes more widely separated afterwards. When the Kozai cycle stops, the mutual inclination between the inner binary and the outer tertiary companion tends to get stuck at a certain value \citep{Fabrycky2007}, making the tertiary companion slightly farther compared to random orientation. However, such difference in projected separations is marginal and cannot explain the steep increase of eclipsing binary fraction at the low-velocity end. Alternatively, the tertiary companions may migrate outwards due to the gravitational perturbations from passing stars, e.g., in the natal open cluster environment \citep{Zakamska2004b}. This may be in line with \cite{El-Badry2019} who suggest such orbital migration to explain the excess of nearly equal-mass binaries with separations of a few thousand AU. However, it implies that nearly all our older short-period binaries should have bright resolved tertiary companions, while we only find $20$\% of our sample having resolved comoving companions with projected separations of $10^{2-5}$\,AU. Therefore, this scenario seems unlikely.  



These alternative scenarios try to explain the increasing eclipsing binary fraction at the low-velocity end. Based the arguments above, they all seem unlikely, and none of them can explain the decreasing eclipsing binary fraction at the high-velocity end. Therefore, we consider the lifetime of short-period binaries to be the best explanation for our results.

\section{Conclusions}
\label{sec:conclusion}

In this paper, we investigate the kinematics of short-period ($<1$\,day) main-sequence eclipsing binaries. We construct two samples of eclipsing binaries: one from the time series analysis of WISE light curves, and the other from the photometric variations in Gaia DR2. These two eclipsing binary samples are complementary to each other: WISE eclipsing binary sample has nearly no contamination, while Gaia eclipsing binary sample has a more homogeneous sky distribution and is not affected by the limitations of period-finding algorithms. We carefully investigate the potential effects from different selection criteria, and require a volume-limited sample instead of magnitude-limited since binaries are brighter than singles. With the kinematics from Gaia DR2, we present the following findings:

\begin{enumerate}
	\item Our primary result is that the eclipsing binary fraction peaks at tangential velocity $V_t=10^{1.3-1.6}$\kms\ and decreases towards both low and high velocity ends (Fig.~\ref{fig:frac-Vt}). 
	\item Since thick-disk and halo stars dominate at high velocity ($V_t>100$\kms), our results imply that the eclipsing binary fraction is at least $\sim10$ times smaller in thick-disk and halo stars compared to thin-disk stars (Fig.~\ref{fig:populations}). This is further supported by the decreasing eclipsing binary fraction when the Galactic latitude $\gtrsim300$\,pc (Fig.~\ref{fig:gal-height}).
	\item The relation between eclipsing binary fraction and kinematics is best explained by the lifetime of eclipsing binaries (Fig.~\ref{fig:model-grid} and \ref{fig:model-accept-reject}). By using Galactic models, we constrain the formation time ($t_0$) of eclipsing binaries to be between 0.6 and 3 Gyr and the disappearing time ($t_1$) to be between 5 and 10\,Gyr, where $t_0$ and $t_1$ are related through $t_0 + 0.4 t_1 \sim5$\,Gyr. The lower eclipsing binary fraction in thick-disk and halo stars may be a consequence of the finite lifetime of eclipsing binaries.
	\item While the pre-MS interaction and the Kozai-Lidov mechanism may help to shrink the binary orbits at an earlier stage, the delayed formation time of $0.6-3$\,Gyr means that short-period binaries cannot form directly from these two scenarios. The timescale is more consistent with magnetic braking, but the upper limit of $\sim3$\,Gyr provides a strict constraint for the theory. 
	\item The disappearance of eclipsing binaries may be due to their mergers within the MS lifetime. This scenario may be tested by studying the kinematics of the merging products, if they can be identified in survey data. 
\end{enumerate}


\acknowledgments

The authors are grateful to the referee, Maxwell Moe, for the constructive report which helped improve the paper significantly. The method to extract variability information from the Gaia DR2 catalog was inspired during 2018 Gaia Data Release 2 Exploration Lab at the European Space Astronomy Centre, where HCH had very useful discussion with A.G.A. Brown, N. Mowlavi, A. Bombrun, L. Palaversa, L. Smith, and E. S. Abrahams. The authors also thank Adam Riess who suggested the investigation of the Galactic models. HCH thanks Smadar Naoz for the insightful discussion on the timescale of the Kozai-Lidov mechanism. HCH would also like to acknowledge helpful conversations with Yuan-Sen Ting, Rosemary Wyse, Sihao Cheng, and Jacob Hamer. HCH was supported by Space@Hopkins and by the Heising-Simons Foundation.

\bibliography{paper-binary_kinematics}{}
\bibliographystyle{aasjournal}

\appendix

\section{Gaia query}

Here is the query for selecting eclipsing binaries from Gaia DR2 used in this paper:

{\obeylines\obeyspaces
	\texttt{
		SELECT
		\ \ gaia.*,
		\ \ allwise.w1mpro, allwise.w2mpro, allwise.w3mpro, allwise.w4mpro,
		\ \ allwise.cc\_flags, allwise.var\_flag
		FROM gaiadr2.gaia\_source AS gaia
		LEFT JOIN gaiadr2.allwise\_best\_neighbour AS allwisexmatch
		\ \ ON gaia.source\_id = allwisexmatch.source\_id
		LEFT JOIN gaiadr1.allwise\_original\_valid AS allwise
		\ \ ON allwise.allwise\_oid = allwisexmatch.allwise\_oid
		WHERE 
		\ \ gaia.parallax\_over\_error > 10 AND 
		\ \ gaia.phot\_g\_mean\_flux\_over\_error>50 AND 
		\ \ gaia.phot\_rp\_mean\_flux\_over\_error>20 AND 
		\ \ gaia.phot\_bp\_mean\_flux\_over\_error>20 AND 
		\ \ gaia.phot\_bp\_rp\_excess\_factor < 1.3+0.06*power(gaia.phot\_bp\_mean\_mag-gaia.phot\_rp\_mean\_mag,2) AND 
		\ \ gaia.phot\_bp\_rp\_excess\_factor > 1.0+0.015*power(gaia.phot\_bp\_mean\_mag-gaia.phot\_rp\_mean\_mag,2) AND 
		\ \ gaia.visibility\_periods\_used>8 AND
		\ \ gaia.parallax >= 2.
}}

\section{Comparison with the Kepler eclipsing binary catalog}
\label{sec:kepler}
In this appendix, we investigate how our eclipsing binary selection, especially in the Gaia-only sample where we do not have light curves, compares with the published Kepler eclipsing binaries. We cross-match the Kepler eclipsing binary catalog \citep{Kirk2016} with Gaia DR2 with a matching radius of 1\,arcsec, and exclude objects that have multiple Gaia matches, ending up with 2721 Kepler eclipsing binaries. The cross-match shows that 95\% of them have parallaxes $<2$\,mas. To compare our MS eclipsing binary selection with the Kepler catalog, we use the following criteria: (1) $0.9<$BP$-$RP$<1.4$, $|\Delta \rm G |< 1.5$, and parallax $>0.5$\,mas, resulting in 665 sources.

\begin{figure}
	\centering
	\includegraphics[width=.5\linewidth]{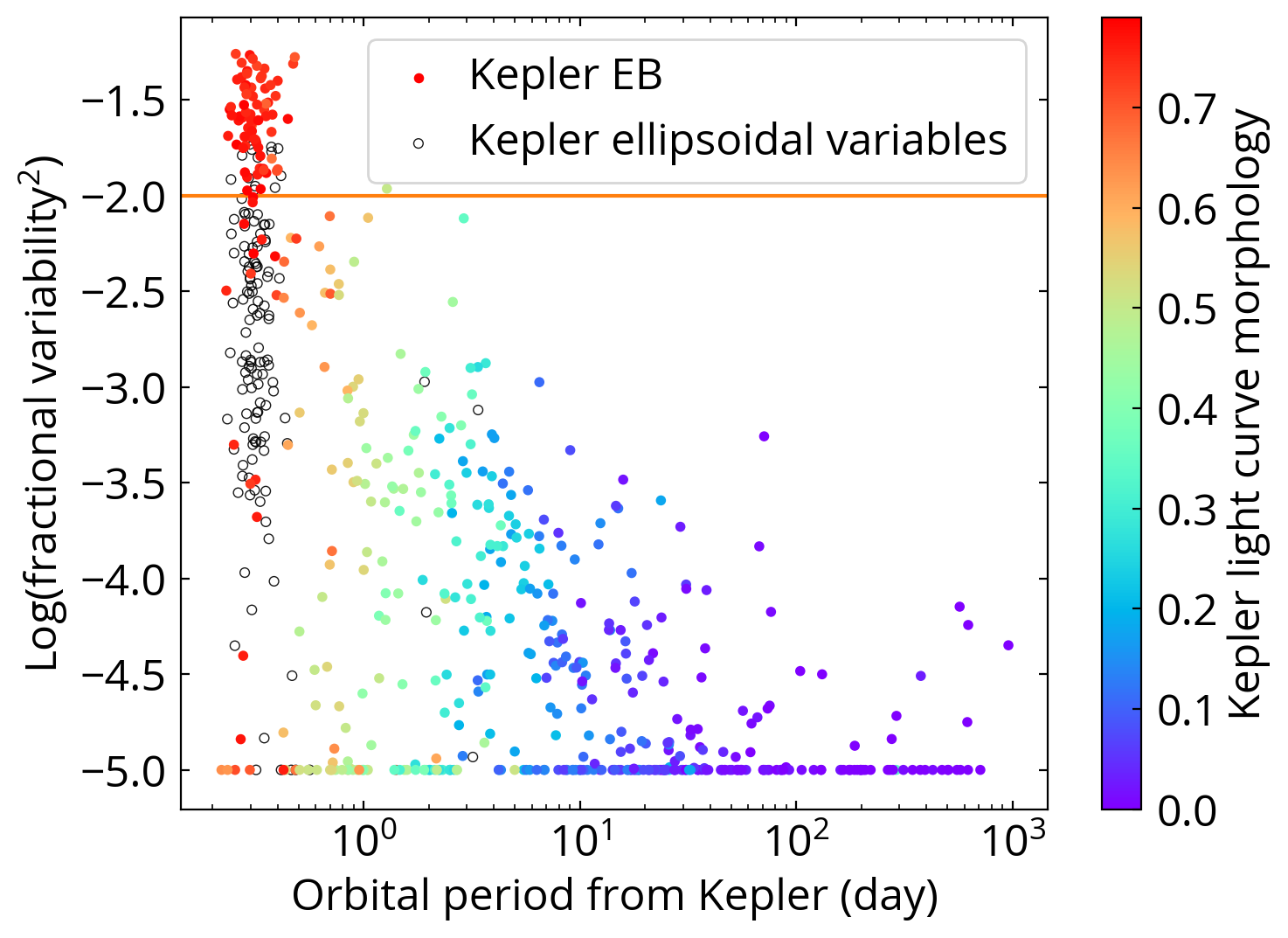}
	\caption{Comparison between fractional variability from Gaia DR2 and Kepler orbital periods. The Kepler sample is divided into eclipsing binaries (solid circles) and ellipsoidal variables (open circles). The solid circles are color-coded by the Kepler light curve morphology measurements \citep{Kirk2016}, and contact binaries usually have values between 0.7 and 0.8. Our Gaia eclipsing binary selection mainly selects contact binaries with orbital periods $<0.5$\,day.}
	\label{fig:KeplerEB-comparison}
\end{figure}

Fig.~\ref{fig:KeplerEB-comparison} shows the Gaia fractional variability versus the orbital periods in the Kepler eclipsing binary catalog. We use the light curve morphology measurements from \cite{Kirk2016} to roughly classify the variables into eclipsing binaries (solid circles) where \texttt{morphology}$<0.8$ and ellipsoidal variables (open circles) where \texttt{morphology}$>0.8$. Eclipsing binaries (solid circles) are color-coded by their \texttt{morphology}, and usually \texttt{morphology}$<0.5$ are detached binaries, \texttt{morphology}$=0.5$-$0.7$ are semi-detached binaries, and \texttt{morphology}$=0.7$-$0.8$ are contact binaries. Our Gaia eclipsing binary selection criterion ($\log(f_G^2) > -2$) is above the orange solid line in the plot. For sources where the Gaia fractional variability is below the instrumental level (Sec.~\ref{sec:gaia-time-series}), we place them at the bottom of the plot ($-5$ on the vertical axis). Most of our Gaia-Kepler eclipsing binaries (85/86) have orbital periods $<0.5$\,day (apparent periods $<0.25$\,day), and most of them are contact binaries where \texttt{morphology}$=0.7$-$0.8$. 

The Kepler eclipsing binary fraction (including ellipsoidal variables) of orbital periods $<1$\,day and cooler K stars (temperatures between 4000-5000 K, similar to our main sample) is $0.4$\%, which is $\sim 4$ times higher than our Gaia EB sample. The difference is due to several factors. (1) First, our Gaia EB selection is more sensitive to contact binaries with orbital periods $<0.5$\,day, and insensitive to (semi-)detached binaries with orbital periods $>0.5$\,day; (2) Second, Kepler is able to recover more low-amplitude ellipsoidal variables at orbital periods $<0.5$\,day. Once we account for these differences, we find that our eclipsing binary fraction is consistent with Kepler.  


It is difficult to compare the WISE eclipsing binary selection with the Kepler eclipsing binary catalog because Kepler eclipsing binaries are more distant and most of them are below or close to WISE's single-exposure sensitivity. However, Fig.~\ref{fig:var-metric} and Fig.~\ref{fig:magpar-cut} have shown that the properties of WISE eclipsing binaries are in general similar to Gaia eclipsing binary selection, so the conclusions from the comparison between the Kepler eclipsing binaries and Gaia eclipsing binaries also apply to the WISE eclipsing binary sample.


\end{document}